# Extended optical waveguide theory with magneto-optical effect and magnetoelectric effect

YOSHIHIRO HONDA,[1,*] ERI IGARASHI,[1] YUYA SHOJI [2] AND TOMOHIRO AMEMIYA[2,*]

[1]*Advanced Research Laboratory, Technology Infrastructure Center, Technology Platform, Sony Group Corporation, 4-14-1 Asahi-cho, Atsugi-shi, Kanagawa 243-0014, Japan*
[2]*Department of Electrical and Electronic Engineering, Tokyo Institute of Technology, Tokyo 152-8522, Japan*
*\*Yoshihiro.Honda@sony.com, amemiya.t.ab@m.titech.ac.jp*

**Abstract:** Optical waveguide theory is essential to the development of various optical devices. Although there are reports on the theory of optical waveguides with magneto-optical (MO) and magnetoelectric (ME) effects, a comprehensive theoretical analysis of waveguides considering these two effects has not yet been published. In this study, the conventional waveguide theory is extended by considering constitutive relations that account for both MO and ME effects. Using the extended waveguide theory, the propagation properties are also analyzed in a medium where metamaterials and magnetic materials are arranged such that MO and ME effects can be controlled independently. It has been confirmed that the interaction between MO and ME effects occurs depending on the arrangement of certain metamaterials and the direction of magnetization. This suggests a nonreciprocal polarization control that rotates the polarization in only one direction when propagating in free space and enhances the nonreciprocal nature of the propagating waves in waveguide propagation.

## 1. Introduction

Optical waveguide theory is an essential theoretical system for the design and development of various optical elements used in today's optical communications. The fundamental elements in optical integrated circuits, such as laser sources, photodetectors, modulators, and transmission lines, all use waveguide geometries. Without optical waveguide theory based on wave equations, these device designs would not be possible. In general, the electric flux density D, magnetic flux density B, electric field E, and magnetic field H can be related as follows:

$$\begin{pmatrix} \mathbf{D} \\ \mathbf{B} \end{pmatrix} = \begin{pmatrix} \bar{\bar{\varepsilon}} & \bar{\bar{\xi}} \\ \bar{\bar{\zeta}} & \bar{\bar{\mu}} \end{pmatrix} \begin{pmatrix} \mathbf{E} \\ \mathbf{H} \end{pmatrix} \qquad (1)$$

where $\bar{\bar{\varepsilon}}, \bar{\bar{\mu}}, \bar{\bar{\xi}}, \bar{\bar{\zeta}}$ are 3 × 3 tensors. In optical waveguide theory, it is important to know how each component of the four tensors interacts with propagating light.

Among the four tensors, the diagonal components, especially $\bar{\bar{\varepsilon}}, \bar{\bar{\mu}}$, represent the dielectric permittivity and magnetic permeability of the material. Optical waveguide theory considering these components is essential for device properties analysis in any material systems, such as compound semiconductor lasers [1–3], passive silicon photonic integrated devices [4–8] optical wavelength filters (multi/demultiplexers) using $SiO_2$ [9–11], and $LiNbO_3$ optical modulators [12–14].

In addition, optical waveguide theory, which takes into account the off-diagonal component of $\bar{\bar{\varepsilon}}$, is mainly used to analyze the nonreciprocal phenomena associated with MO effects. The Faraday effect, the Cotton-Mouton effect, and the magneto-optical Kerr effect are well-known nonreciprocal phenomena [15]. These effects are caused by the off-diagonal component of $\bar{\bar{\varepsilon}}$ being changed by the external magnetic field. Optical waveguide theory, which accounts for the off-diagonal component of $\bar{\bar{\varepsilon}}$ is essential for analyzing the properties of garnet-based optical isolators/circulators [16] used to suppress reflected light in fiber optic communications and waveguide-based optical isolators [17–20] that can be integrated into optical integrated circuits.

Of the four tensors in Eq. (1), the optical waveguide theory, which considers $\bar{\bar{\xi}}$ and $\bar{\bar{\zeta}}$ is mainly used to calculate the polarization state associated with ME effects. In the past, it was used to analyze the propagation of circularly polarized light in chiral media [21,22]. More recently, it has been used to analyze bi-isotropic materials such as multiferroic materials and metamaterials when incorporated in bulk media and waveguides [23–26].

As mentioned earlier, many research groups have reported on free space and waveguide propagation considering MO effects [27–29], and ME effects [30,31] (only in certain exceptional cases for waveguide propagation). However, a systematic theory for optical waveguides that also considers MO and ME effects has not yet been proposed and discussed. Therefore, in this study, we extended the conventional wave equation to include the constitutive relations that account for all four tensors $\bar{\bar{\varepsilon}}, \bar{\bar{\mu}}, \bar{\bar{\xi}}, \bar{\bar{\zeta}}$ in Eq. (1), and developed a systematic optical waveguide theory.

The remainder of this study is structured as follows. First, in Section 2, constitutive relations involving both MO and ME effects are considered, and refer to specific examples to realize them. In Section 3, the conventional wave equation is extended by using it to derive a systematic optical waveguide theory. In Section 4, the derived optical waveguide theory is applied to free space and waveguide propagation and some typical examples are discussed (some of the results are attributed to the previously known equations). Finally, Section 5 discusses unique optical propagation properties in systems where both MO and ME effects are considered, with specific analytical results for both free space and waveguide propagation.

It is expected that the results described in this study will be versatile in the design of optical elements with different functional materials and very useful for the development of new wave control elements.

## 2. Constitutive relations

In this section, the constitutive relations of the material are mentioned as a preliminary step to derive the extended wave equations considering MO and ME effects. First, **D** and **B** are generally related by the following relations.

$$\mathbf{D} = \bar{\bar{\varepsilon}}\mathbf{E} + \bar{\bar{\xi}}\mathbf{H} \qquad (2)$$

$$\mathbf{B} = \bar{\bar{\zeta}}\mathbf{E} + \bar{\bar{\mu}}\mathbf{H} \qquad (3)$$

where $\bar{\bar{\varepsilon}}$, $\bar{\bar{\mu}}$, $\bar{\bar{\xi}}$ and $\bar{\bar{\zeta}}$ are the permittivity, permeability, magnetic-to-electric coupling, and electric-to-magnetic coupling tensors, each with 3 × 3 dimensions, respectively. In a general medium, the electric and magnetic coupling is bidirectional and is expressed as $\bar{\bar{\zeta}} = -\bar{\bar{\xi}}^T$ [32]. Therefore, considering MO and ME effects, the electric and magnetic flux densities are determined in the following equation.

$$\begin{pmatrix} D_x \\ D_y \\ D_z \\ B_x \\ B_y \\ B_z \end{pmatrix} = \begin{pmatrix} \begin{pmatrix} \varepsilon & a & -c \\ -a & \varepsilon & b \\ c & -b & \varepsilon \end{pmatrix} & \begin{pmatrix} G & A & D \\ E & H & B \\ C & F & I \end{pmatrix} \\ -\begin{pmatrix} G & E & C \\ A & H & F \\ D & B & I \end{pmatrix} & \begin{pmatrix} \mu & 0 & 0 \\ 0 & \mu & 0 \\ 0 & 0 & \mu \end{pmatrix} \end{pmatrix} \begin{pmatrix} E_x \\ E_y \\ E_z \\ H_x \\ H_y \\ H_z \end{pmatrix} \qquad (4)$$

In Eq. (4), the tensor components are as follows: $\varepsilon$ and $\mu$ are the parameters corresponding to the diagonal components of the permittivity and permeability tensors of the medium. For simplicity, an isotropic medium is assumed, and all these values are equal.

$a$, $b$, and $c$ are the parameters corresponding to the off-diagonal components of the permittivity tensor of the medium and take values only when the medium has ferromagnetism. For example, if the magnetization of a ferromagnetic medium is aligned in a particular direction, as shown in Fig. 1, $a$, $b$, and $c$ have complex values. Various isolators and circulators have been realized by integrating ferromagnetic materials such as Ce:YIG, Fe, Co, and Ni into waveguides. These devices are based on the principle that the off-diagonal component of the permittivity tensor causes nonreciprocal effects [33–36].

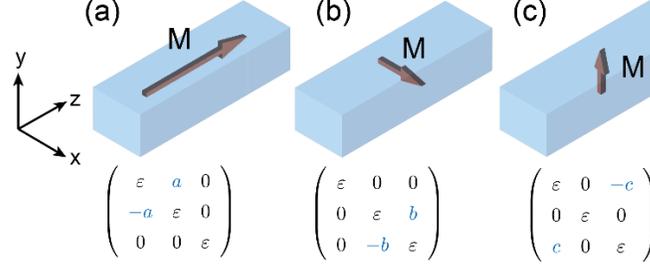

**Fig. 1.** Three configurations of magneto-optical effect.

$A - I$ are the components of the magnetic-to-electric coupling tensor (electric-to-magnetic coupling tensor), which take values only when the medium has bi-anisotropic properties. For example, when multiferroic materials such as $Cr_2O_3$, $TbMnO_3$, etc. are used, the off-diagonal components $A - F$ have complex values. In other ways, with appropriately designed metamaterials all components, including the diagonal components $G$, $H$, and $I$, can be controlled arbitrarily. For example, Fig. 2 shows split ring resonator (SRR) and helical metamaterial arrangements yielding each component of $A - I$. The direction of the incident electric field and the induced magnetic field determines which tensor component has a value. Slow light propagation and compact optical modulators have been realized by integrating SRR-based metamaterials into waveguides. These take advantage of the steep refractive index dispersion or the change in refractive index due to $A - I$ having a value of [37,38].

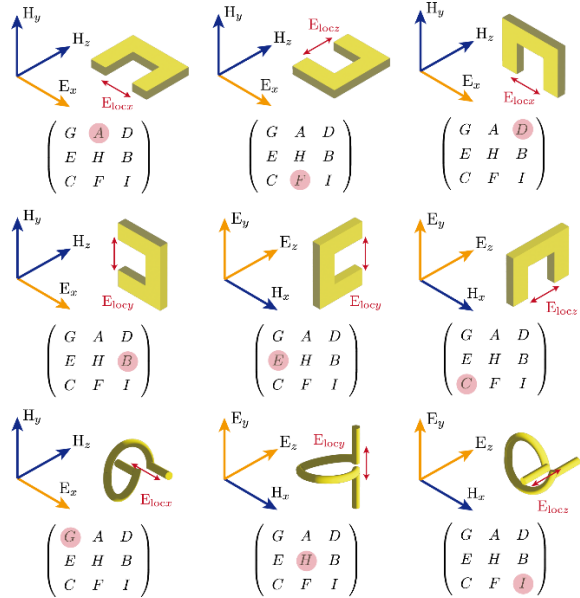

**Fig. 2.** Nine orientations of the SRR and helical metamaterials with respect to the polarization of incident light.

For a summary of the above, see Fig. 3. For example, to create MO and ME effects simultaneously ($a - c$ and $A - I$ have simultaneous values), ferromagnetic materials and metamaterials can be integrated [39], as shown in Fig. 4. In this case, the mode and polarization state of the propagating light become more complicated. However, with the extended wave equation derived in this study, it is possible to perform these propagation analyses uniformly.

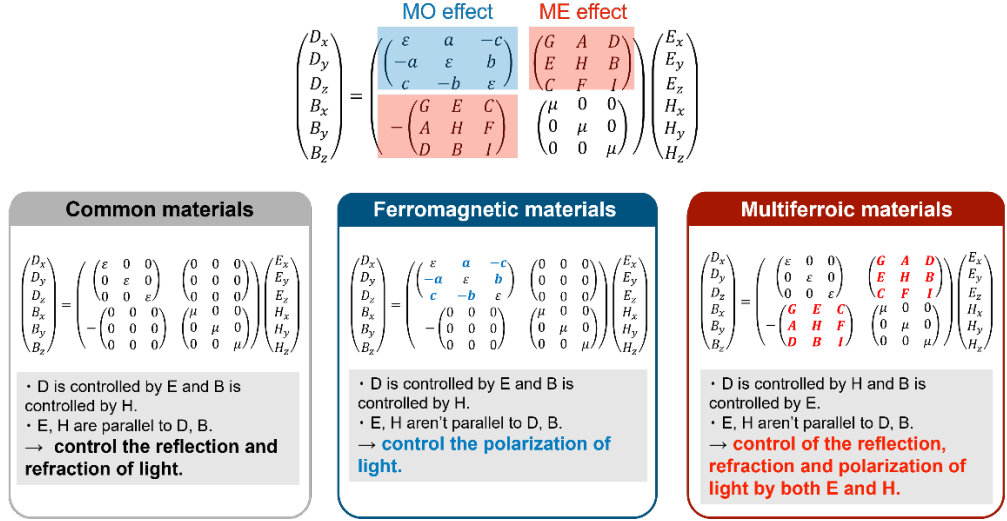

**Fig. 3**. Classification of the material according to MO effect and ME effect

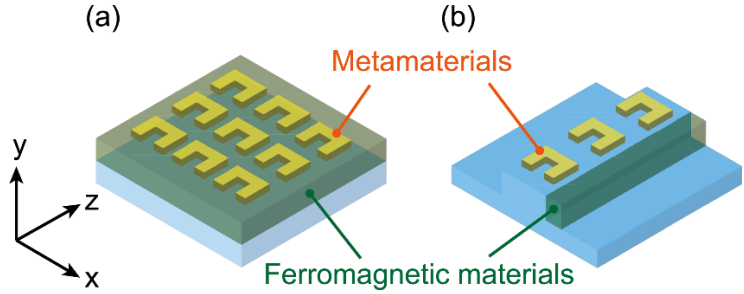

**Fig. 4.** (a) Bulk ferromagnetic material with metamaterials. (b) Metamaterial waveguide with ferromagnetic layer.

## 3. Extended optical waveguide theory with MO and ME effects

The constitutive relations in Eq. (4) are used to derive the extended wave equations that include both MO and ME effects. Before describing the details, the flow of the derivation is shown in Fig. 5.

### 3.1 Derivation of the extended wave equations

We begin with Maxwell's equations.

$$i\omega D = -\nabla \times H \tag{5}$$
$$i\omega B = \nabla \times E \tag{6}$$

Considering these equations in matrix form, we obtain the following equation, where $\omega$ denotes the angular frequency.

$$i\omega\begin{pmatrix}D_x\\D_y\\D_z\\B_x\\B_y\\B_z\end{pmatrix}=\begin{pmatrix}0 & -\begin{pmatrix}0 & -\partial_z & \partial_y\\ \partial_z & 0 & -\partial_x\\ -\partial_y & \partial_x & 0\end{pmatrix}\\ \begin{pmatrix}0 & -\partial_z & \partial_y\\ \partial_z & 0 & -\partial_x\\ -\partial_y & \partial_x & 0\end{pmatrix} & 0\end{pmatrix}\begin{pmatrix}E_x\\E_y\\E_z\\H_x\\H_y\\H_z\end{pmatrix} \quad (7)$$

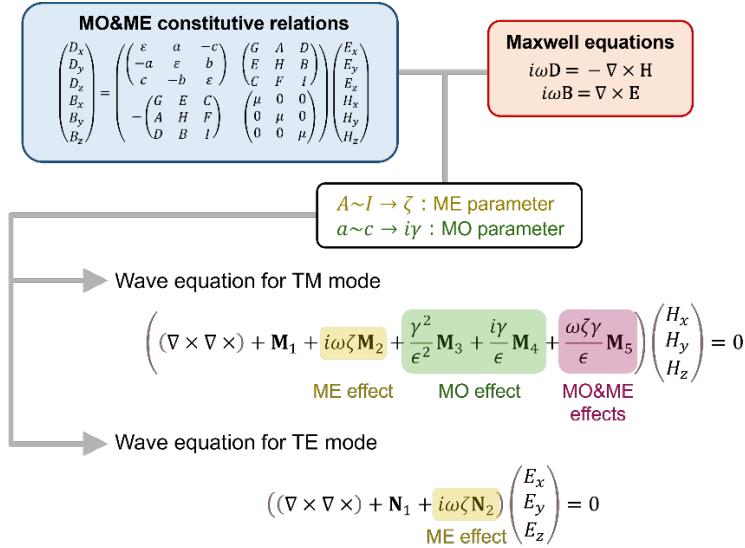

**Fig. 5.** Flowchart of the introduction of the wave equations for MO and ME effects.

From Eq. (4) and Eq. (7), we obtain

$$i\omega\begin{pmatrix}\begin{pmatrix}\varepsilon & a & -c\\ -a & \varepsilon & b\\ c & -b & \varepsilon\end{pmatrix} & \begin{pmatrix}G & A & D\\ E & H & B\\ C & F & I\end{pmatrix}\\ -\begin{pmatrix}G & E & C\\ A & H & F\\ D & B & I\end{pmatrix} & \begin{pmatrix}\mu & 0 & 0\\ 0 & \mu & 0\\ 0 & 0 & \mu\end{pmatrix}\end{pmatrix}\begin{pmatrix}E_x\\E_y\\E_z\\H_x\\H_y\\H_z\end{pmatrix}=\begin{pmatrix}0 & -\begin{pmatrix}0 & -\partial_z & \partial_y\\ \partial_z & 0 & -\partial_x\\ -\partial_y & \partial_x & 0\end{pmatrix}\\ \begin{pmatrix}0 & -\partial_z & \partial_y\\ \partial_z & 0 & -\partial_x\\ -\partial_y & \partial_x & 0\end{pmatrix} & 0\end{pmatrix}\begin{pmatrix}E_x\\E_y\\E_z\\H_x\\H_y\\H_z\end{pmatrix} \quad (8)$$

For the above equation, each line is enumerated as follows

$$i\omega(\varepsilon E_x + aE_y - cE_z + GH_x + AH_y + DH_z) = \partial_z H_y - \partial_y H_z \quad (9a)$$
$$i\omega(-aE_x + \varepsilon E_y + bE_z + EH_x + HH_y + BH_z) = -\partial_z H_x + \partial_x H_z \quad (9b)$$
$$i\omega(cE_x - bE_y + \varepsilon E_z + CH_x + FH_y + IH_z) = \partial_y H_x - \partial_x H_y \quad (9c)$$
$$i\omega(-GE_x - EE_y - CE_z + \mu H_x) = -\partial_z E_y + \partial_y E_z \quad (10a)$$
$$i\omega(-AE_x - HE_y - FE_z + \mu H_y) = \partial_z E_x - \partial_x E_z \quad (10b)$$
$$i\omega(-DE_x - BE_y - IE_z + \mu H_z) = -\partial_y E_x + \partial_x E_y \quad (10c)$$

When deriving the wave equation with these equations, it is easier to deal with the assumption of a certain fundamental mode. Therefore, in this study, the wave equations are derived for the TM and TE modes, which are the most common in waveguide optics. The extended wave equations for the TM and TE modes are obtained by the following procedure using Eqs. (9a)-(9c) and (10a)-(10c).

TM mode: Eqs. (9a)-(9c) are solved for the electric field and then substitute them into Eqs. (10a)-(10c).

TE mode: Eqs. (10a)-(10c) are solved for the magnetic field and then substitute them into Eqs. (9a)-(9c).

Depending on the value of each tensor component $a - c$ and $A - I$, there are exceptional cases (see Section 4 for details). For example, the TM and TE modes interact in a complex way in a waveguide. Moreover, completely different polarization states, such as left circular polarization (LCP) and right circular polarization (RCP), become the fundamental modes. Such cases can be easily analyzed starting from the extended wave equations for the TM and TE modes.

### 3.2 Extended wave equation for TM modes

As mentioned in the previous section, we can obtain the wave equations for the TM mode by substituting them into Eqs. (9a)-(9c) after solving Eqs. (10a)-(10c) for the electric field. For simplicity, we assume that $i\gamma$ enters $a - c$, the element for MO effect, and $\zeta$ enters $A - I$, the element for ME effect. The extended wave equation for the TM mode, which includes both MO and ME effects, is formulated as follows.

$$\left( (\nabla \times \nabla \times) + \mathbf{M}_1 + i\omega\zeta \mathbf{M}_2 + \frac{\gamma^2}{\epsilon^2}\mathbf{M}_3 + \frac{i\gamma}{\epsilon}\mathbf{M}_4 + \frac{\omega\zeta\gamma}{\epsilon}\mathbf{M}_5 \right)\begin{pmatrix} H_x \\ H_y \\ H_z \end{pmatrix} = 0 \tag{11a}$$

$$\begin{pmatrix} E_x \\ E_y \\ E_z \end{pmatrix} = \mathbf{Z}_1^{-1}\mathbf{Z}_2\begin{pmatrix} H_x \\ H_y \\ H_z \end{pmatrix} \tag{11b}$$

where $\mathbf{M}_1$-$\mathbf{M}_5$ and $\mathbf{Z}_1$ and $\mathbf{Z}_2$ are all in tensor form. In particular, the coefficients of $\mathbf{M}_2$, $\mathbf{M}_3$, $\mathbf{M}_4$, and $\mathbf{M}_5$ are $\zeta$ related to ME effect, $\gamma^2$ and $\gamma$ related to MO effect, and $\zeta\gamma$, which represents the interaction between MO and ME effects. Thus, $\mathbf{M}_2$-$\mathbf{M}_5$ are tensors that mainly represent these effects. Here, $\mathbf{M}_1$ is called the wave propagation tensor, $\mathbf{M}_2$ is called ME tensor, $\mathbf{M}_3$ is called the second-order MO tensor, $\mathbf{M}_4$ is called the first-order MO tensor, $\mathbf{M}_5$ is called the MO and ME tensor, $\mathbf{Z}_1$ and $\mathbf{Z}_2$ are called the impedance tensor.

In Eqs. (11a) and (11b), a situation is assumed where one of $a - c$ and one of $A - I$ have values. All results are summarized for 27 types in total ($3 \times 9$), and their forms are listed in Table 1 (see Supplementary for the derivation of each tensor component). When $i\gamma$ enters all elements of $a - c$ and $\zeta$ enters all elements of $A - I$, $\mathbf{M}_1$ to $\mathbf{M}_5$ and $\mathbf{Z}_1$, $\mathbf{Z}_2$ are given as follows.

$$\mathbf{M}_1 = -\omega^2\varepsilon\mu\left(1 - \frac{\gamma^2}{\epsilon^2}\right)\mathbf{I} - 3\omega^2\zeta^2\left(1 - \frac{\gamma^2}{\epsilon^2}\right)\mathbf{I} \tag{12a}$$

$$\mathbf{M}_2 = \left(1 - \frac{\gamma^2}{\epsilon^2}\right)\left(2(\nabla \times) + \begin{pmatrix} 0 & \partial_x + \partial_y & -\partial_x - \partial_z \\ -\partial_x - \partial_y & 0 & \partial_y + \partial_z \\ \partial_x + \partial_z & -\partial_y - \partial_z & 0 \end{pmatrix}\right) \tag{12b}$$

$$\mathbf{M}_3 = -(\nabla \times \nabla \times) \tag{12c}$$

$$\mathbf{M}_4 = (\partial_x + \partial_y + \partial_z)(\nabla \times) \tag{12d}$$

$$\mathbf{M}_5 = 0 \tag{12e}$$

$$\mathbf{Z}_1 = \begin{pmatrix} 1 & \dfrac{i\gamma}{\epsilon} & -\dfrac{i\gamma}{\epsilon} \\ -\dfrac{i\gamma}{\epsilon} & 1 & \dfrac{i\gamma}{\epsilon} \\ \dfrac{i\gamma}{\epsilon} & -\dfrac{i\gamma}{\epsilon} & 1 \end{pmatrix} \tag{12f}$$

$$\mathbf{Z}_2 = -\dfrac{1}{i\omega\varepsilon}(\nabla \times) - \dfrac{\zeta}{\varepsilon}\begin{pmatrix} 1 & 1 & 1 \\ 1 & 1 & 1 \\ 1 & 1 & 1 \end{pmatrix} \tag{12g}$$

**Table 1. Summary of tensor parameters for the TM mode**

| Selected parameter pair | $\mathbf{M}_1$ | $\mathbf{M}_2$ | $\mathbf{M}_3$ | $\mathbf{M}_4$ | $\mathbf{M}_5$ | $\mathbf{Z}_1$ | $\mathbf{Z}_2$ |
|---|---|---|---|---|---|---|---|
| A-a | $-\omega^2\varepsilon\mu\left(1-\dfrac{\gamma^2}{\epsilon^2}\right)\mathbf{I}$ $-\omega^2\zeta^2\begin{pmatrix}0&0&0\\0&1&0\\0&0&0\end{pmatrix}$ | $\begin{pmatrix}0&0&0\\0&0&\partial_y\\0&-\partial_y&0\end{pmatrix}$ | $\begin{pmatrix}\partial_y\partial_y&-\partial_x\partial_y&0\\-\partial_x\partial_y&\partial_x\partial_x&0\\0&0&0\end{pmatrix}$ | $-\partial_z(\nabla\times)$ | $\begin{pmatrix}0&\partial_z&0\\\partial_z&0&-\partial_x\\0&-\partial_x&0\end{pmatrix}$ | $\begin{pmatrix}1&\dfrac{i\gamma}{\epsilon}&0\\-\dfrac{i\gamma}{\epsilon}&1&0\\0&0&1\end{pmatrix}$ | $-\dfrac{1}{i\omega\varepsilon}(\nabla\times)$ $-\dfrac{\zeta}{\varepsilon}\begin{pmatrix}0&1&0\\0&0&0\\0&0&0\end{pmatrix}$ |
| A-b | $-\omega^2\varepsilon\mu\left(1-\dfrac{\gamma^2}{\epsilon^2}\right)\mathbf{I}$ $-\omega^2\zeta^2\begin{pmatrix}1\end{pmatrix}$ $-\dfrac{\gamma^2}{\epsilon^2}\begin{pmatrix}0&0&0\\0&1&0\\0&0&0\end{pmatrix}$ | $\begin{pmatrix}1\end{pmatrix}$ $-\dfrac{\gamma^2}{\epsilon^2}\begin{pmatrix}0&0&0\\0&0&\partial_y\\0&-\partial_y&0\end{pmatrix}$ | $\begin{pmatrix}0&0&0\\0&\partial_z\partial_z&-\partial_y\partial_z\\0&-\partial_y\partial_z&\partial_y\partial_y\end{pmatrix}$ | $-\partial_x(\nabla\times)$ | $0$ | $\begin{pmatrix}1&0&0\\0&1&\dfrac{i\gamma}{\epsilon}\\0&-\dfrac{i\gamma}{\epsilon}&1\end{pmatrix}$ | $-\dfrac{1}{i\omega\varepsilon}(\nabla\times)$ $-\dfrac{\zeta}{\varepsilon}\begin{pmatrix}0&1&0\\0&0&0\\0&0&0\end{pmatrix}$ |
| A-c | $-\omega^2\varepsilon\mu\left(1-\dfrac{\gamma^2}{\epsilon^2}\right)\mathbf{I}$ $-\omega^2\zeta^2\begin{pmatrix}0&0&0\\0&1&0\\0&0&0\end{pmatrix}$ | $\begin{pmatrix}0&0&0\\0&0&\partial_y\\0&-\partial_y&0\end{pmatrix}$ | $\begin{pmatrix}\partial_z\partial_z&0&-\partial_x\partial_z\\0&0&0\\-\partial_x\partial_z&0&\partial_x\partial_x\end{pmatrix}$ | $-\partial_y(\nabla\times)$ | $\begin{pmatrix}0&\partial_y&0\\\partial_y&-2\partial_x&0\\0&0&0\end{pmatrix}$ | $\begin{pmatrix}1&0&-\dfrac{i\gamma}{\epsilon}\\0&1&0\\\dfrac{i\gamma}{\epsilon}&0&1\end{pmatrix}$ | $-\dfrac{1}{i\omega\varepsilon}(\nabla\times)$ $-\dfrac{\zeta}{\varepsilon}\begin{pmatrix}0&1&0\\0&0&0\\0&0&0\end{pmatrix}$ |
| B-a | $-\omega^2\varepsilon\mu\left(1-\dfrac{\gamma^2}{\epsilon^2}\right)\mathbf{I}$ $-\omega^2\zeta^2\begin{pmatrix}0&0&0\\0&0&0\\0&0&1\end{pmatrix}$ | $\begin{pmatrix}0&0&-\partial_z\\0&0&0\\\partial_z&0&0\end{pmatrix}$ | $\begin{pmatrix}\partial_y\partial_y&-\partial_x\partial_y&0\\-\partial_x\partial_y&\partial_x\partial_x&0\\0&0&0\end{pmatrix}$ | $-\partial_z(\nabla\times)$ | $\begin{pmatrix}0&0&0\\0&0&\partial_z\\0&\partial_z&-2\partial_y\end{pmatrix}$ | $\begin{pmatrix}1&\dfrac{i\gamma}{\epsilon}&0\\-\dfrac{i\gamma}{\epsilon}&1&0\\0&0&1\end{pmatrix}$ | $-\dfrac{1}{i\omega\varepsilon}(\nabla\times)$ $-\dfrac{\zeta}{\varepsilon}\begin{pmatrix}0&0&0\\0&0&1\\0&0&0\end{pmatrix}$ |
| B-b | $-\omega^2\varepsilon\mu\left(1-\dfrac{\gamma^2}{\epsilon^2}\right)\mathbf{I}$ $-\omega^2\zeta^2\begin{pmatrix}0&0&0\\0&0&0\\0&0&1\end{pmatrix}$ | $\begin{pmatrix}0&0&-\partial_z\\0&0&0\\\partial_z&0&0\end{pmatrix}$ | $\begin{pmatrix}0&0&0\\0&\partial_z\partial_z&-\partial_y\partial_z\\0&-\partial_y\partial_z&\partial_y\partial_y\end{pmatrix}$ | $-\partial_x(\nabla\times)$ | $\begin{pmatrix}0&0&-\partial_y\\0&0&\partial_x\\-\partial_y&\partial_x&0\end{pmatrix}$ | $\begin{pmatrix}1&0&0\\0&1&\dfrac{i\gamma}{\epsilon}\\0&-\dfrac{i\gamma}{\epsilon}&1\end{pmatrix}$ | $-\dfrac{1}{i\omega\varepsilon}(\nabla\times)$ $-\dfrac{\zeta}{\varepsilon}\begin{pmatrix}0&0&0\\0&0&1\\0&0&0\end{pmatrix}$ |
| B-c | $-\omega^2\varepsilon\mu\left(1-\dfrac{\gamma^2}{\epsilon^2}\right)\mathbf{I}$ $-\omega^2\zeta^2\begin{pmatrix}1\end{pmatrix}$ $-\dfrac{\gamma^2}{\epsilon^2}\begin{pmatrix}0&0&0\\0&0&0\\0&0&1\end{pmatrix}$ | $\begin{pmatrix}1\end{pmatrix}$ $-\dfrac{\gamma^2}{\epsilon^2}\begin{pmatrix}0&0&-\partial_z\\0&0&0\\\partial_z&0&0\end{pmatrix}$ | $\begin{pmatrix}\partial_z\partial_z&0&-\partial_x\partial_z\\0&0&0\\-\partial_x\partial_z&0&\partial_x\partial_x\end{pmatrix}$ | $-\partial_y(\nabla\times)$ | $0$ | $\begin{pmatrix}1&0&-\dfrac{i\gamma}{\epsilon}\\0&1&0\\\dfrac{i\gamma}{\epsilon}&0&1\end{pmatrix}$ | $-\dfrac{1}{i\omega\varepsilon}(\nabla\times)$ $-\dfrac{\zeta}{\varepsilon}\begin{pmatrix}0&0&0\\0&0&1\\0&0&0\end{pmatrix}$ |
| C-a | $-\omega^2\varepsilon\mu\left(1-\dfrac{\gamma^2}{\epsilon^2}\right)\mathbf{I}$ $-\omega^2\zeta^2\begin{pmatrix}1\end{pmatrix}$ $-\dfrac{\gamma^2}{\epsilon^2}\begin{pmatrix}1&0&0\\0&0&0\\0&0&0\end{pmatrix}$ | $\begin{pmatrix}1\end{pmatrix}$ $-\dfrac{\gamma^2}{\epsilon^2}\begin{pmatrix}0&\partial_x&0\\-\partial_x&0&0\\0&0&0\end{pmatrix}$ | $\begin{pmatrix}\partial_y\partial_y&-\partial_x\partial_y&0\\-\partial_x\partial_y&\partial_x\partial_x&0\\0&0&0\end{pmatrix}$ | $-\partial_z(\nabla\times)$ | $0$ | $\begin{pmatrix}1&\dfrac{i\gamma}{\epsilon}&0\\-\dfrac{i\gamma}{\epsilon}&1&0\\0&0&1\end{pmatrix}$ | $-\dfrac{1}{i\omega\varepsilon}(\nabla\times)$ $-\dfrac{\zeta}{\varepsilon}\begin{pmatrix}0&0&0\\0&0&0\\1&0&0\end{pmatrix}$ |
| C-b | $-\omega^2\varepsilon\mu\left(1-\dfrac{\gamma^2}{\epsilon^2}\right)\mathbf{I}$ $-\omega^2\zeta^2\begin{pmatrix}1&0&0\\0&0&0\\0&0&0\end{pmatrix}$ | $\begin{pmatrix}0&\partial_x&0\\-\partial_x&0&0\\0&0&0\end{pmatrix}$ | $\begin{pmatrix}0&0&0\\0&\partial_z\partial_z&-\partial_y\partial_z\\0&-\partial_y\partial_z&\partial_y\partial_y\end{pmatrix}$ | $-\partial_x(\nabla\times)$ | $\begin{pmatrix}-2\partial_z&0&\partial_x\\0&0&0\\\partial_x&0&0\end{pmatrix}$ | $\begin{pmatrix}1&0&0\\0&1&\dfrac{i\gamma}{\epsilon}\\0&-\dfrac{i\gamma}{\epsilon}&1\end{pmatrix}$ | $-\dfrac{1}{i\omega\varepsilon}(\nabla\times)$ $-\dfrac{\zeta}{\varepsilon}\begin{pmatrix}0&0&0\\0&0&0\\1&0&0\end{pmatrix}$ |
| C-c | $-\omega^2\varepsilon\mu\left(1-\dfrac{\gamma^2}{\epsilon^2}\right)\mathbf{I}$ $-\omega^2\zeta^2\begin{pmatrix}1&0&0\\0&0&0\\0&0&0\end{pmatrix}$ | $\begin{pmatrix}0&\partial_x&0\\-\partial_x&0&0\\0&0&0\end{pmatrix}$ | $\begin{pmatrix}\partial_z\partial_z&0&-\partial_x\partial_z\\0&0&0\\-\partial_x\partial_z&0&\partial_x\partial_x\end{pmatrix}$ | $-\partial_y(\nabla\times)$ | $\begin{pmatrix}0&-\partial_z&\partial_y\\-\partial_z&0&0\\\partial_y&0&0\end{pmatrix}$ | $\begin{pmatrix}1&0&-\dfrac{i\gamma}{\epsilon}\\0&1&0\\\dfrac{i\gamma}{\epsilon}&0&1\end{pmatrix}$ | $-\dfrac{1}{i\omega\varepsilon}(\nabla\times)$ $-\dfrac{\zeta}{\varepsilon}\begin{pmatrix}0&0&0\\0&0&0\\1&0&0\end{pmatrix}$ |
| D-a | $-\omega^2\varepsilon\mu\left(1-\dfrac{\gamma^2}{\epsilon^2}\right)\mathbf{I}$ $-\omega^2\zeta^2\begin{pmatrix}0&0&0\\0&0&0\\0&0&1\end{pmatrix}$ | $\begin{pmatrix}0&0&0\\0&0&\partial_z\\0&-\partial_z&0\end{pmatrix}$ | $\begin{pmatrix}\partial_y\partial_y&-\partial_x\partial_y&0\\-\partial_x\partial_y&\partial_x\partial_x&0\\0&0&0\end{pmatrix}$ | $-\partial_z(\nabla\times)$ | $\begin{pmatrix}0&0&\partial_z\\0&0&0\\\partial_z&0&-2\partial_x\end{pmatrix}$ | $\begin{pmatrix}1&\dfrac{i\gamma}{\epsilon}&0\\-\dfrac{i\gamma}{\epsilon}&1&0\\0&0&1\end{pmatrix}$ | $-\dfrac{1}{i\omega\varepsilon}(\nabla\times)$ $-\dfrac{\zeta}{\varepsilon}\begin{pmatrix}0&0&1\\0&0&0\\0&0&0\end{pmatrix}$ |
| D-b | $-\omega^2\varepsilon\mu\left(1-\dfrac{\gamma^2}{\epsilon^2}\right)\mathbf{I}$ $-\omega^2\zeta^2\begin{pmatrix}1\end{pmatrix}$ $-\dfrac{\gamma^2}{\epsilon^2}\begin{pmatrix}0&0&0\\0&0&0\\0&0&1\end{pmatrix}$ | $\begin{pmatrix}1\end{pmatrix}$ $-\dfrac{\gamma^2}{\epsilon^2}\begin{pmatrix}0&0&0\\0&0&\partial_z\\0&-\partial_z&0\end{pmatrix}$ | $\begin{pmatrix}0&0&0\\0&\partial_z\partial_z&-\partial_y\partial_z\\0&-\partial_y\partial_z&\partial_y\partial_y\end{pmatrix}$ | $-\partial_x(\nabla\times)$ | $0$ | $\begin{pmatrix}1&0&0\\0&1&\dfrac{i\gamma}{\epsilon}\\0&-\dfrac{i\gamma}{\epsilon}&1\end{pmatrix}$ | $-\dfrac{1}{i\omega\varepsilon}(\nabla\times)$ $-\dfrac{\zeta}{\varepsilon}\begin{pmatrix}0&0&1\\0&0&0\\0&0&0\end{pmatrix}$ |
| D-c | $-\omega^2\varepsilon\mu\left(1-\frac{\gamma^2}{\epsilon^2}\right)\mathbf{I}$ $-\omega^2\zeta^2\begin{pmatrix}0&0&0\\0&0&0\\0&0&1\end{pmatrix}$ | $\begin{pmatrix}0&0&0\\0&0&\partial_z\\0&-\partial_z&0\end{pmatrix}$ | $\begin{pmatrix}\partial_z\partial_z&0&-\partial_x\partial_z\\0&0&0\\-\partial_x\partial_z&0&\partial_x\partial_x\end{pmatrix}$ | $-\partial_y(\nabla\times)$ | $\begin{pmatrix}0&0&\partial_y\\0&0&-\partial_x\\\partial_y&-\partial_x&0\end{pmatrix}$ | $\begin{pmatrix}1&0&-\frac{i\gamma}{\epsilon}\\0&1&0\\\frac{i\gamma}{\epsilon}&0&1\end{pmatrix}$ | $-\frac{1}{i\omega\varepsilon}(\nabla\times)$ $-\frac{\zeta}{\varepsilon}\begin{pmatrix}0&0&1\\0&0&0\\0&0&0\end{pmatrix}$ |
| E-a | $-\omega^2\varepsilon\mu\left(1-\frac{\gamma^2}{\epsilon^2}\right)\mathbf{I}$ $-\omega^2\zeta^2\begin{pmatrix}1&0&0\\0&0&0\\0&0&0\end{pmatrix}$ | $\begin{pmatrix}0&0&-\partial_x\\0&0&0\\\partial_x&0&0\end{pmatrix}$ | $\begin{pmatrix}\partial_y\partial_y&-\partial_x\partial_y&0\\-\partial_x\partial_y&\partial_x\partial_x&0\\0&0&0\end{pmatrix}$ | $-\partial_z(\nabla\times)$ | $\begin{pmatrix}0&\partial_z&-\partial_y\\\partial_z&0&0\\-\partial_y&0&0\end{pmatrix}$ | $\begin{pmatrix}1&\frac{i\gamma}{\epsilon}&0\\-\frac{i\gamma}{\epsilon}&1&0\\0&0&1\end{pmatrix}$ | $-\frac{1}{i\omega\varepsilon}(\nabla\times)$ $-\frac{\zeta}{\varepsilon}\begin{pmatrix}0&0&0\\1&0&0\\0&0&0\end{pmatrix}$ |
| E-b | $-\omega^2\varepsilon\mu\left(1-\frac{\gamma^2}{\epsilon^2}\right)\mathbf{I}$ $-\omega^2\zeta^2\begin{pmatrix}1&0&0\\0&0&0\\0&0&0\end{pmatrix}$ | $\begin{pmatrix}0&0&-\partial_x\\0&0&0\\\partial_x&0&0\end{pmatrix}$ | $\begin{pmatrix}0&0&0\\0&\partial_z\partial_z&-\partial_y\partial_z\\0&-\partial_y\partial_z&\partial_y\partial_y\end{pmatrix}$ | $-\partial_x(\nabla\times)$ | $\begin{pmatrix}-2\partial_x&\partial_x&0\\\partial_x&0&0\\0&0&0\end{pmatrix}$ | $\begin{pmatrix}1&0&0\\0&1&\frac{i\gamma}{\epsilon}\\0&-\frac{i\gamma}{\epsilon}&1\end{pmatrix}$ | $-\frac{1}{i\omega\varepsilon}(\nabla\times)$ $-\frac{\zeta}{\varepsilon}\begin{pmatrix}0&0&0\\1&0&0\\0&0&0\end{pmatrix}$ |
| E-c | $-\omega^2\varepsilon\mu\left(1-\frac{\gamma^2}{\epsilon^2}\right)\mathbf{I}$ $-\omega^2\zeta^2\left(1-\frac{\gamma^2}{\epsilon^2}\right)\begin{pmatrix}1&0&0\\0&0&0\\0&0&0\end{pmatrix}$ | $\left(1-\frac{\gamma^2}{\epsilon^2}\right)\begin{pmatrix}0&0&-\partial_x\\0&0&0\\\partial_x&0&0\end{pmatrix}$ | $\begin{pmatrix}\partial_z\partial_z&0&-\partial_x\partial_z\\0&0&0\\-\partial_x\partial_z&0&\partial_x\partial_x\end{pmatrix}$ | $-\partial_y(\nabla\times)$ | $0$ | $\begin{pmatrix}1&0&-\frac{i\gamma}{\epsilon}\\0&1&0\\\frac{i\gamma}{\epsilon}&0&1\end{pmatrix}$ | $-\frac{1}{i\omega\varepsilon}(\nabla\times)$ $-\frac{\zeta}{\varepsilon}\begin{pmatrix}0&0&0\\1&0&0\\0&0&0\end{pmatrix}$ |
| F-a | $-\omega^2\varepsilon\mu\left(1-\frac{\gamma^2}{\epsilon^2}\right)\mathbf{I}$ $-\omega^2\zeta^2\left(1-\frac{\gamma^2}{\epsilon^2}\right)\begin{pmatrix}0&0&0\\0&1&0\\0&0&0\end{pmatrix}$ | $\left(1-\frac{\gamma^2}{\epsilon^2}\right)\begin{pmatrix}0&\partial_y&0\\-\partial_y&0&0\\0&0&0\end{pmatrix}$ | $\begin{pmatrix}\partial_y\partial_y&-\partial_x\partial_y&0\\-\partial_x\partial_y&\partial_x\partial_x&0\\0&0&0\end{pmatrix}$ | $-\partial_z(\nabla\times)$ | $0$ | $\begin{pmatrix}1&\frac{i\gamma}{\epsilon}&0\\-\frac{i\gamma}{\epsilon}&1&0\\0&0&1\end{pmatrix}$ | $-\frac{1}{i\omega\varepsilon}(\nabla\times)$ $-\frac{\zeta}{\varepsilon}\begin{pmatrix}0&0&0\\0&0&0\\0&1&0\end{pmatrix}$ |
| F-b | $-\omega^2\varepsilon\mu\left(1-\frac{\gamma^2}{\epsilon^2}\right)\mathbf{I}$ $-\omega^2\zeta^2\begin{pmatrix}0&0&0\\0&1&0\\0&0&0\end{pmatrix}$ | $\begin{pmatrix}0&\partial_y&0\\-\partial_y&0&0\\0&0&0\end{pmatrix}$ | $\begin{pmatrix}0&0&0\\0&\partial_z\partial_z&-\partial_y\partial_z\\0&-\partial_y\partial_z&\partial_y\partial_y\end{pmatrix}$ | $-\partial_x(\nabla\times)$ | $\begin{pmatrix}0&-\partial_z&0\\-\partial_z&0&\partial_x\\0&\partial_x&0\end{pmatrix}$ | $\begin{pmatrix}1&0&0\\0&1&\frac{i\gamma}{\epsilon}\\0&-\frac{i\gamma}{\epsilon}&1\end{pmatrix}$ | $-\frac{1}{i\omega\varepsilon}(\nabla\times)$ $-\frac{\zeta}{\varepsilon}\begin{pmatrix}0&0&0\\0&0&0\\0&1&0\end{pmatrix}$ |
| F-c | $-\omega^2\varepsilon\mu\left(1-\frac{\gamma^2}{\epsilon^2}\right)\mathbf{I}$ $-\omega^2\zeta^2\begin{pmatrix}0&0&0\\0&1&0\\0&0&0\end{pmatrix}$ | $\begin{pmatrix}0&\partial_y&0\\-\partial_y&0&0\\0&0&0\end{pmatrix}$ | $\begin{pmatrix}\partial_z\partial_z&0&-\partial_x\partial_z\\0&0&0\\-\partial_x\partial_z&0&\partial_x\partial_x\end{pmatrix}$ | $-\partial_y(\nabla\times)$ | $\begin{pmatrix}0&0&0\\0&-2\partial_z&\partial_y\\0&\partial_y&0\end{pmatrix}$ | $\begin{pmatrix}1&0&-\frac{i\gamma}{\epsilon}\\0&1&0\\\frac{i\gamma}{\epsilon}&0&1\end{pmatrix}$ | $-\frac{1}{i\omega\varepsilon}(\nabla\times)$ $-\frac{\zeta}{\varepsilon}\begin{pmatrix}0&0&0\\0&0&0\\0&1&0\end{pmatrix}$ |
| G-a | $-\omega^2\varepsilon\mu\left(1-\frac{\gamma^2}{\epsilon^2}\right)\mathbf{I}$ $-\omega^2\zeta^2\begin{pmatrix}1&0&0\\0&0&0\\0&0&0\end{pmatrix}$ | $(\nabla\times)\big|_{\partial_x=0}$ | $\begin{pmatrix}\partial_y\partial_y&-\partial_x\partial_y&0\\-\partial_x\partial_y&\partial_x\partial_x&0\\0&0&0\end{pmatrix}$ | $-\partial_z(\nabla\times)$ | $\begin{pmatrix}2\partial_z&-\partial_x&0\\0&0&0\\-\partial_x&0&0\end{pmatrix}$ | $\begin{pmatrix}1&\frac{i\gamma}{\epsilon}&0\\-\frac{i\gamma}{\epsilon}&1&0\\0&0&1\end{pmatrix}$ | $-\frac{1}{i\omega\varepsilon}(\nabla\times)$ $-\frac{\zeta}{\varepsilon}\begin{pmatrix}1&0&0\\0&0&0\\0&0&0\end{pmatrix}$ |
| G-b | $-\omega^2\varepsilon\mu\left(1-\frac{\gamma^2}{\epsilon^2}\right)\mathbf{I}$ $-\omega^2\zeta^2\left(1-\frac{\gamma^2}{\epsilon^2}\right)\begin{pmatrix}1&0&0\\0&0&0\\0&0&0\end{pmatrix}$ | $\left(1-\frac{\gamma^2}{\epsilon^2}\right)(\nabla\times)\big|_{\partial_x=0}$ | $\begin{pmatrix}0&0&0\\0&\partial_z\partial_z&-\partial_y\partial_z\\0&-\partial_y\partial_z&\partial_y\partial_y\end{pmatrix}$ | $-\partial_x(\nabla\times)$ | $0$ | $\begin{pmatrix}1&0&0\\0&1&\frac{i\gamma}{\epsilon}\\0&-\frac{i\gamma}{\epsilon}&1\end{pmatrix}$ | $-\frac{1}{i\omega\varepsilon}(\nabla\times)$ $-\frac{\zeta}{\varepsilon}\begin{pmatrix}1&0&0\\0&0&0\\0&0&0\end{pmatrix}$ |
| G-c | $-\omega^2\varepsilon\mu\left(1-\frac{\gamma^2}{\epsilon^2}\right)\mathbf{I}$ $-\omega^2\zeta^2\begin{pmatrix}1&0&0\\0&0&0\\0&0&0\end{pmatrix}$ | $(\nabla\times)\big|_{\partial_x=0}$ | $\begin{pmatrix}\partial_z\partial_z&0&-\partial_x\partial_z\\0&0&0\\-\partial_x\partial_z&0&\partial_x\partial_x\end{pmatrix}$ | $-\partial_y(\nabla\times)$ | $\begin{pmatrix}2\partial_y&-\partial_x&0\\-\partial_x&0&0\\0&0&0\end{pmatrix}$ | $\begin{pmatrix}1&0&-\frac{i\gamma}{\epsilon}\\0&1&0\\\frac{i\gamma}{\epsilon}&0&1\end{pmatrix}$ | $-\frac{1}{i\omega\varepsilon}(\nabla\times)$ $-\frac{\zeta}{\varepsilon}\begin{pmatrix}1&0&0\\0&0&0\\0&0&0\end{pmatrix}$ |
| H-a | $-\omega^2\varepsilon\mu\left(1-\frac{\gamma^2}{\epsilon^2}\right)\mathbf{I}$ $-\omega^2\zeta^2\begin{pmatrix}0&0&0\\0&1&0\\0&0&0\end{pmatrix}$ | $(\nabla\times)\big|_{\partial_y=0}$ | $\begin{pmatrix}\partial_y\partial_y&-\partial_x\partial_y&0\\-\partial_x\partial_y&\partial_x\partial_x&0\\0&0&0\end{pmatrix}$ | $-\partial_z(\nabla\times)$ | $\begin{pmatrix}0&0&0\\0&2\partial_z&-\partial_y\\0&-\partial_y&0\end{pmatrix}$ | $\begin{pmatrix}1&\frac{i\gamma}{\epsilon}&0\\-\frac{i\gamma}{\epsilon}&1&0\\0&0&1\end{pmatrix}$ | $-\frac{1}{i\omega\varepsilon}(\nabla\times)$ $-\frac{\zeta}{\varepsilon}\begin{pmatrix}0&0&0\\0&1&0\\0&0&0\end{pmatrix}$ |
| H-b | $-\omega^2\varepsilon\mu\left(1-\frac{\gamma^2}{\epsilon^2}\right)\mathbf{I}$ $-\omega^2\zeta^2\begin{pmatrix}0&0&0\\0&1&0\\0&0&0\end{pmatrix}$ | $(\nabla\times)\big|_{\partial_y=0}$ | $\begin{pmatrix}0&0&0\\0&\partial_z\partial_z&-\partial_y\partial_z\\0&-\partial_y\partial_z&\partial_y\partial_y\end{pmatrix}$ | $-\partial_x(\nabla\times)$ | $\begin{pmatrix}0&-\partial_y&0\\-\partial_y&2\partial_x&0\\0&0&0\end{pmatrix}$ | $\begin{pmatrix}1&0&0\\0&1&\frac{i\gamma}{\epsilon}\\0&-\frac{i\gamma}{\epsilon}&1\end{pmatrix}$ | $-\frac{1}{i\omega\varepsilon}(\nabla\times)$ $-\frac{\zeta}{\varepsilon}\begin{pmatrix}0&0&0\\0&1&0\\0&0&0\end{pmatrix}$ |
| H-c | $-\omega^2\varepsilon\mu\left(1-\frac{\gamma^2}{\epsilon^2}\right)\mathbf{I}$ $-\omega^2\zeta^2\left(1-\frac{\gamma^2}{\epsilon^2}\right)\begin{pmatrix}0&0&0\\0&1&0\\0&0&0\end{pmatrix}$ | $\left(1-\frac{\gamma^2}{\epsilon^2}\right)(\nabla\times)\big|_{\partial_y=0}$ | $\begin{pmatrix}\partial_z\partial_z&0&-\partial_x\partial_z\\0&0&0\\-\partial_x\partial_z&0&\partial_x\partial_x\end{pmatrix}$ | $-\partial_y(\nabla\times)$ | $0$ | $\begin{pmatrix}1&0&-\frac{i\gamma}{\epsilon}\\0&1&0\\\frac{i\gamma}{\epsilon}&0&1\end{pmatrix}$ | $-\frac{1}{i\omega\varepsilon}(\nabla\times)$ $-\frac{\zeta}{\varepsilon}\begin{pmatrix}0&0&0\\0&1&0\\0&0&0\end{pmatrix}$ |
| I-a | $-\omega^2\varepsilon\mu\left(1-\frac{\gamma^2}{\epsilon^2}\right)\mathbf{I}$ $-\omega^2\zeta^2\left(1-\frac{\gamma^2}{\epsilon^2}\right)\begin{pmatrix}0&0&0\\0&0&0\\0&0&1\end{pmatrix}$ | $\left(1-\frac{\gamma^2}{\epsilon^2}\right)(\nabla\times)\big|_{\partial_z=0}$ | $\begin{pmatrix}\partial_y\partial_y&-\partial_x\partial_y&0\\-\partial_x\partial_y&\partial_x\partial_x&0\\0&0&0\end{pmatrix}$ | $-\partial_z(\nabla\times)$ | $0$ | $\begin{pmatrix}1&\frac{i\gamma}{\epsilon}&0\\-\frac{i\gamma}{\epsilon}&1&0\\0&0&1\end{pmatrix}$ | $-\frac{1}{i\omega\varepsilon}(\nabla\times)$ $-\frac{\zeta}{\varepsilon}\begin{pmatrix}0&0&0\\0&0&0\\0&0&1\end{pmatrix}$ |
| I-b | $-\omega^2\varepsilon\mu\left(1-\frac{\gamma^2}{\epsilon^2}\right)\mathbf{I}$ $-\omega^2\zeta^2\begin{pmatrix}0&0&0\\0&0&0\\0&0&1\end{pmatrix}$ | $(\nabla\times)\big|_{\partial_z=0}$ | $\begin{pmatrix}0&0&0\\0&\partial_z\partial_z&-\partial_y\partial_z\\0&-\partial_y\partial_z&\partial_y\partial_y\end{pmatrix}$ | $-\partial_x(\nabla\times)$ | $\begin{pmatrix}0&0&-\partial_z\\0&0&0\\-\partial_z&0&2\partial_x\end{pmatrix}$ | $\begin{pmatrix}1&0&0\\0&1&\frac{i\gamma}{\epsilon}\\0&-\frac{i\gamma}{\epsilon}&1\end{pmatrix}$ | $-\frac{1}{i\omega\varepsilon}(\nabla\times)$ $-\frac{\zeta}{\varepsilon}\begin{pmatrix}0&0&0\\0&0&0\\0&0&1\end{pmatrix}$ |
| I-c | $-\omega^2\varepsilon\mu\left(1-\frac{\gamma^2}{\epsilon^2}\right)\mathbf{I}$ $-\omega^2\zeta^2\begin{pmatrix}0&0&0\\0&0&0\\0&0&1\end{pmatrix}$ | $(\nabla\times)\big|_{\partial_z=0}$ | $\begin{pmatrix}\partial_z\partial_z&0&-\partial_x\partial_z\\0&0&0\\-\partial_x\partial_z&0&\partial_x\partial_x\end{pmatrix}$ | $-\partial_y(\nabla\times)$ | $\begin{pmatrix}0&0&0\\0&0&-\partial_z\\0&-\partial_z&2\partial_y\end{pmatrix}$ | $\begin{pmatrix}1&0&-\frac{i\gamma}{\epsilon}\\0&1&0\\\frac{i\gamma}{\epsilon}&0&1\end{pmatrix}$ | $-\frac{1}{i\omega\varepsilon}(\nabla\times)$ $-\frac{\zeta}{\varepsilon}\begin{pmatrix}0&0&0\\0&0&0\\0&0&1\end{pmatrix}$ |

## 3.2 Extended wave equation for TE modes

After rewriting Eqs. (9a)-(9c) for the magnetic field as described in Section 3.1, we can obtain the wave equation for the TE mode by substituting them into Eqs. (8a)-(8c). As in Section 3.2, if $a - c$, the elements for MO effect, are $i\gamma$ and $A - I$, the elements for ME effect, are $\zeta$, then the extended wave equation for the TE mode, which includes both MO and ME effects, are formulated as follows.

$$\left((\nabla \times \nabla \times) + \mathbf{N}_1 + i\omega\zeta \mathbf{N}_2\right) \begin{pmatrix} E_x \\ E_y \\ E_z \end{pmatrix} = 0 \tag{13a}$$

$$\begin{pmatrix} H_x \\ H_y \\ H_z \end{pmatrix} = \mathbf{Y} \begin{pmatrix} E_x \\ E_y \\ E_z \end{pmatrix} \tag{13b}$$

where $\mathbf{N}_1$, $\mathbf{N}_2$ and $\mathbf{Y}$ are all in tensor form. In particular, the coefficients of $\mathbf{N}_2$ is $\zeta$ related to ME effect. Therefore, in this study, $\mathbf{N}_1$ will be referred to as the wave propagation tensor, $\mathbf{N}_2$ as the ME tensor and $\mathbf{Y}$ as the admittance tensor.

In Eqs. (13a) and (13b), we assume a situation where one instance of $a - c$ and one instance of $A - I$ have values. All results are summarized for 27 types in total ($3 \times 9$), and their forms in Table 2 (see Supplementary for the derivation of each tensor component). When $i\gamma$ enters all elements of $a - c$ and $\zeta$ enters all elements of $A - I$, $\mathbf{N}_1$, $\mathbf{N}_2$ and $\mathbf{Y}$ are given as follows.

$$\mathbf{N}_1 = -\omega^2 \mu\varepsilon \mathbf{I} - \omega^2 \mu \begin{pmatrix} 0 & -i\gamma & i\gamma \\ i\gamma & 0 & -i\gamma \\ -i\gamma & i\gamma & 0 \end{pmatrix} - 3\omega^2 \zeta^2 \mathbf{I} \tag{14a}$$

$$\mathbf{N}_2 = 2(\nabla \times) + \begin{pmatrix} 0 & \partial_x + \partial_y & -\partial_x - \partial_z \\ -\partial_x - \partial_y & 0 & \partial_y + \partial_z \\ \partial_x + \partial_z & -\partial_y - \partial_z & 0 \end{pmatrix} \tag{14b}$$

$$\mathbf{Y} = \frac{1}{i\omega\mu}(\nabla \times) + \frac{\zeta}{\mu}\begin{pmatrix} 1 & 1 & 1 \\ 1 & 1 & 1 \\ 1 & 1 & 1 \end{pmatrix} \tag{14c}$$

**Table 2. Summary of tensor parameters for the TE mode**

| | | Each tensor form | | |
|---|---|---|---|---|
| | | $\mathbf{N}_1$ | $\mathbf{N}_2$ | $\mathbf{Y}$ |
| Selected parameter pair | A-a | $-\omega^2\mu\varepsilon\mathbf{I} - \omega^2\mu\begin{pmatrix}0 & i\gamma & 0 \\ -i\gamma & 0 & 0 \\ 0 & 0 & 0\end{pmatrix} - \omega^2\zeta^2\begin{pmatrix}1 & 0 & 0 \\ 0 & 0 & 0 \\ 0 & 0 & 0\end{pmatrix}$ | $\begin{pmatrix}0 & 0 & -\partial_x \\ 0 & 0 & 0 \\ \partial_x & 0 & 0\end{pmatrix}$ | $\frac{1}{i\omega\mu}(\nabla\times) + \frac{\zeta}{\mu}\begin{pmatrix}0 & 0 & 0 \\ 1 & 0 & 0 \\ 0 & 0 & 0\end{pmatrix}$ |
| | A-b | $-\omega^2\mu\varepsilon\mathbf{I} - \omega^2\mu\begin{pmatrix}0 & 0 & 0 \\ 0 & 0 & i\gamma \\ 0 & -i\gamma & 0\end{pmatrix} - \omega^2\zeta^2\begin{pmatrix}1 & 0 & 0 \\ 0 & 0 & 0 \\ 0 & 0 & 0\end{pmatrix}$ | $\begin{pmatrix}0 & 0 & -\partial_x \\ 0 & 0 & 0 \\ \partial_x & 0 & 0\end{pmatrix}$ | $\frac{1}{i\omega\mu}(\nabla\times) + \frac{\zeta}{\mu}\begin{pmatrix}0 & 0 & 0 \\ 1 & 0 & 0 \\ 0 & 0 & 0\end{pmatrix}$ |
| | A-c | $-\omega^2\mu\varepsilon\mathbf{I} - \omega^2\mu\begin{pmatrix}0 & 0 & -i\gamma \\ 0 & 0 & 0 \\ i\gamma & 0 & 0\end{pmatrix} - \omega^2\zeta^2\begin{pmatrix}1 & 0 & 0 \\ 0 & 0 & 0 \\ 0 & 0 & 0\end{pmatrix}$ | $\begin{pmatrix}0 & 0 & -\partial_x \\ 0 & 0 & 0 \\ \partial_x & 0 & 0\end{pmatrix}$ | $\frac{1}{i\omega\mu}(\nabla\times) + \frac{\zeta}{\mu}\begin{pmatrix}0 & 0 & 0 \\ 1 & 0 & 0 \\ 0 & 0 & 0\end{pmatrix}$ |
| | B-a | $-\omega^2\mu\varepsilon\mathbf{I} - \omega^2\mu\begin{pmatrix}0 & i\gamma & 0 \\ -i\gamma & 0 & 0 \\ 0 & 0 & 0\end{pmatrix} - \omega^2\zeta^2\begin{pmatrix}0 & 0 & 0 \\ 0 & 1 & 0 \\ 0 & 0 & 0\end{pmatrix}$ | $\begin{pmatrix}0 & \partial_y & 0 \\ -\partial_y & 0 & 0 \\ 0 & 0 & 0\end{pmatrix}$ | $\frac{1}{i\omega\mu}(\nabla\times) + \frac{\zeta}{\mu}\begin{pmatrix}0 & 0 & 0 \\ 0 & 0 & 0 \\ 0 & 1 & 0\end{pmatrix}$ |
| | B-b | $-\omega^2\mu\varepsilon\mathbf{I} - \omega^2\mu\begin{pmatrix}0 & 0 & 0 \\ 0 & 0 & i\gamma \\ 0 & -i\gamma & 0\end{pmatrix} - \omega^2\zeta^2\begin{pmatrix}0 & 0 & 0 \\ 0 & 1 & 0 \\ 0 & 0 & 0\end{pmatrix}$ | $\begin{pmatrix}0 & \partial_y & 0 \\ -\partial_y & 0 & 0 \\ 0 & 0 & 0\end{pmatrix}$ | $\frac{1}{i\omega\mu}(\nabla\times) + \frac{\zeta}{\mu}\begin{pmatrix}0 & 0 & 0 \\ 0 & 0 & 0 \\ 0 & 1 & 0\end{pmatrix}$ |
| B-c | $-\omega^2\mu\varepsilon\,\mathbf{I} - \omega^2\mu\begin{pmatrix}0 & 0 & -i\gamma\\ 0 & 0 & 0\\ i\gamma & 0 & 0\end{pmatrix} - \omega^2\zeta^2\begin{pmatrix}0 & 0 & 0\\ 0 & 1 & 0\\ 0 & 0 & 0\end{pmatrix}$ | $\begin{pmatrix}0 & \partial_y & 0\\ -\partial_y & 0 & 0\\ 0 & 0 & 0\end{pmatrix}$ | $\dfrac{1}{i\omega\mu}(\nabla\times) + \dfrac{\zeta}{\mu}\begin{pmatrix}0 & 0 & 0\\ 0 & 0 & 0\\ 0 & 1 & 0\end{pmatrix}$ |
| C-a | $-\omega^2\mu\varepsilon\,\mathbf{I} - \omega^2\mu\begin{pmatrix}0 & i\gamma & 0\\ -i\gamma & 0 & 0\\ 0 & 0 & 0\end{pmatrix} - \omega^2\zeta^2\begin{pmatrix}0 & 0 & 0\\ 0 & 0 & 0\\ 0 & 0 & 1\end{pmatrix}$ | $\begin{pmatrix}0 & 0 & 0\\ 0 & 0 & \partial_z\\ 0 & -\partial_z & 0\end{pmatrix}$ | $\dfrac{1}{i\omega\mu}(\nabla\times) + \dfrac{\zeta}{\mu}\begin{pmatrix}0 & 0 & 1\\ 0 & 0 & 0\\ 0 & 0 & 0\end{pmatrix}$ |
| C-b | $-\omega^2\mu\varepsilon\,\mathbf{I} - \omega^2\mu\begin{pmatrix}0 & 0 & 0\\ 0 & 0 & i\gamma\\ 0 & -i\gamma & 0\end{pmatrix} - \omega^2\zeta^2\begin{pmatrix}0 & 0 & 0\\ 0 & 0 & 0\\ 0 & 0 & 1\end{pmatrix}$ | $\begin{pmatrix}0 & 0 & 0\\ 0 & 0 & \partial_z\\ 0 & -\partial_z & 0\end{pmatrix}$ | $\dfrac{1}{i\omega\mu}(\nabla\times) + \dfrac{\zeta}{\mu}\begin{pmatrix}0 & 0 & 1\\ 0 & 0 & 0\\ 0 & 0 & 0\end{pmatrix}$ |
| C-c | $-\omega^2\mu\varepsilon\,\mathbf{I} - \omega^2\mu\begin{pmatrix}0 & 0 & -i\gamma\\ 0 & 0 & 0\\ i\gamma & 0 & 0\end{pmatrix} - \omega^2\zeta^2\begin{pmatrix}0 & 0 & 0\\ 0 & 0 & 0\\ 0 & 0 & 1\end{pmatrix}$ | $\begin{pmatrix}0 & 0 & 0\\ 0 & 0 & \partial_z\\ 0 & -\partial_z & 0\end{pmatrix}$ | $\dfrac{1}{i\omega\mu}(\nabla\times) + \dfrac{\zeta}{\mu}\begin{pmatrix}0 & 0 & 1\\ 0 & 0 & 0\\ 0 & 0 & 0\end{pmatrix}$ |
| D-a | $-\omega^2\mu\varepsilon\,\mathbf{I} - \omega^2\mu\begin{pmatrix}0 & i\gamma & 0\\ -i\gamma & 0 & 0\\ 0 & 0 & 0\end{pmatrix} - \omega^2\zeta^2\begin{pmatrix}1 & 0 & 0\\ 0 & 0 & 0\\ 0 & 0 & 0\end{pmatrix}$ | $\begin{pmatrix}0 & \partial_x & 0\\ -\partial_x & 0 & 0\\ 0 & 0 & 0\end{pmatrix}$ | $\dfrac{1}{i\omega\mu}(\nabla\times) + \dfrac{\zeta}{\mu}\begin{pmatrix}0 & 0 & 0\\ 0 & 0 & 0\\ 1 & 0 & 0\end{pmatrix}$ |
| D-b | $-\omega^2\mu\varepsilon\,\mathbf{I} - \omega^2\mu\begin{pmatrix}0 & 0 & 0\\ 0 & 0 & i\gamma\\ 0 & -i\gamma & 0\end{pmatrix} - \omega^2\zeta^2\begin{pmatrix}1 & 0 & 0\\ 0 & 0 & 0\\ 0 & 0 & 0\end{pmatrix}$ | $\begin{pmatrix}0 & \partial_x & 0\\ -\partial_x & 0 & 0\\ 0 & 0 & 0\end{pmatrix}$ | $\dfrac{1}{i\omega\mu}(\nabla\times) + \dfrac{\zeta}{\mu}\begin{pmatrix}0 & 0 & 0\\ 0 & 0 & 0\\ 1 & 0 & 0\end{pmatrix}$ |
| D-c | $-\omega^2\mu\varepsilon\,\mathbf{I} - \omega^2\mu\begin{pmatrix}0 & 0 & -i\gamma\\ 0 & 0 & 0\\ i\gamma & 0 & 0\end{pmatrix} - \omega^2\zeta^2\begin{pmatrix}1 & 0 & 0\\ 0 & 0 & 0\\ 0 & 0 & 0\end{pmatrix}$ | $\begin{pmatrix}0 & \partial_x & 0\\ -\partial_x & 0 & 0\\ 0 & 0 & 0\end{pmatrix}$ | $\dfrac{1}{i\omega\mu}(\nabla\times) + \dfrac{\zeta}{\mu}\begin{pmatrix}0 & 0 & 0\\ 0 & 0 & 0\\ 1 & 0 & 0\end{pmatrix}$ |
| E-a | $-\omega^2\mu\varepsilon\,\mathbf{I} - \omega^2\mu\begin{pmatrix}0 & i\gamma & 0\\ -i\gamma & 0 & 0\\ 0 & 0 & 0\end{pmatrix} - \omega^2\zeta^2\begin{pmatrix}0 & 0 & 0\\ 0 & 1 & 0\\ 0 & 0 & 0\end{pmatrix}$ | $\begin{pmatrix}0 & 0 & 0\\ 0 & 0 & \partial_y\\ 0 & -\partial_y & 0\end{pmatrix}$ | $\dfrac{1}{i\omega\mu}(\nabla\times) + \dfrac{\zeta}{\mu}\begin{pmatrix}0 & 1 & 0\\ 0 & 0 & 0\\ 0 & 0 & 0\end{pmatrix}$ |
| E-b | $-\omega^2\mu\varepsilon\,\mathbf{I} - \omega^2\mu\begin{pmatrix}0 & 0 & 0\\ 0 & 0 & i\gamma\\ 0 & -i\gamma & 0\end{pmatrix} - \omega^2\zeta^2\begin{pmatrix}0 & 0 & 0\\ 0 & 1 & 0\\ 0 & 0 & 0\end{pmatrix}$ | $\begin{pmatrix}0 & 0 & 0\\ 0 & 0 & \partial_y\\ 0 & -\partial_y & 0\end{pmatrix}$ | $\dfrac{1}{i\omega\mu}(\nabla\times) + \dfrac{\zeta}{\mu}\begin{pmatrix}0 & 1 & 0\\ 0 & 0 & 0\\ 0 & 0 & 0\end{pmatrix}$ |
| E-c | $-\omega^2\mu\varepsilon\,\mathbf{I} - \omega^2\mu\begin{pmatrix}0 & 0 & -i\gamma\\ 0 & 0 & 0\\ i\gamma & 0 & 0\end{pmatrix} - \omega^2\zeta^2\begin{pmatrix}0 & 0 & 0\\ 0 & 1 & 0\\ 0 & 0 & 0\end{pmatrix}$ | $\begin{pmatrix}0 & 0 & 0\\ 0 & 0 & \partial_y\\ 0 & -\partial_y & 0\end{pmatrix}$ | $\dfrac{1}{i\omega\mu}(\nabla\times) + \dfrac{\zeta}{\mu}\begin{pmatrix}0 & 1 & 0\\ 0 & 0 & 0\\ 0 & 0 & 0\end{pmatrix}$ |
| F-a | $-\omega^2\mu\varepsilon\,\mathbf{I} - \omega^2\mu\begin{pmatrix}0 & i\gamma & 0\\ -i\gamma & 0 & 0\\ 0 & 0 & 0\end{pmatrix} - \omega^2\zeta^2\begin{pmatrix}0 & 0 & 0\\ 0 & 0 & 0\\ 0 & 0 & 1\end{pmatrix}$ | $\begin{pmatrix}0 & 0 & -\partial_z\\ 0 & 0 & 0\\ \partial_z & 0 & 0\end{pmatrix}$ | $\dfrac{1}{i\omega\mu}(\nabla\times) + \dfrac{\zeta}{\mu}\begin{pmatrix}0 & 0 & 0\\ 0 & 0 & 1\\ 0 & 0 & 0\end{pmatrix}$ |
| F-b | $-\omega^2\mu\varepsilon\,\mathbf{I} - \omega^2\mu\begin{pmatrix}0 & 0 & 0\\ 0 & 0 & i\gamma\\ 0 & -i\gamma & 0\end{pmatrix} - \omega^2\zeta^2\begin{pmatrix}0 & 0 & 0\\ 0 & 0 & 0\\ 0 & 0 & 1\end{pmatrix}$ | $\begin{pmatrix}0 & 0 & -\partial_z\\ 0 & 0 & 0\\ \partial_z & 0 & 0\end{pmatrix}$ | $\dfrac{1}{i\omega\mu}(\nabla\times) + \dfrac{\zeta}{\mu}\begin{pmatrix}0 & 0 & 0\\ 0 & 0 & 1\\ 0 & 0 & 0\end{pmatrix}$ |
| F-c | $-\omega^2\mu\varepsilon\,\mathbf{I} - \omega^2\mu\begin{pmatrix}0 & 0 & -i\gamma\\ 0 & 0 & 0\\ i\gamma & 0 & 0\end{pmatrix} - \omega^2\zeta^2\begin{pmatrix}0 & 0 & 0\\ 0 & 0 & 0\\ 0 & 0 & 1\end{pmatrix}$ | $\begin{pmatrix}0 & 0 & -\partial_z\\ 0 & 0 & 0\\ \partial_z & 0 & 0\end{pmatrix}$ | $\dfrac{1}{i\omega\mu}(\nabla\times) + \dfrac{\zeta}{\mu}\begin{pmatrix}0 & 0 & 0\\ 0 & 0 & 1\\ 0 & 0 & 0\end{pmatrix}$ |
| G-a | $-\omega^2\mu\varepsilon\,\mathbf{I} - \omega^2\mu\begin{pmatrix}0 & i\gamma & 0\\ -i\gamma & 0 & 0\\ 0 & 0 & 0\end{pmatrix} - \omega^2\zeta^2\begin{pmatrix}1 & 0 & 0\\ 0 & 0 & 0\\ 0 & 0 & 0\end{pmatrix}$ | $(\nabla\times)\vert_{\partial_x=0}$ | $\dfrac{1}{i\omega\mu}(\nabla\times) + \dfrac{\zeta}{\mu}\begin{pmatrix}1 & 0 & 0\\ 0 & 0 & 0\\ 0 & 0 & 0\end{pmatrix}$ |
| G-b | $-\omega^2\mu\varepsilon\,\mathbf{I} - \omega^2\mu\begin{pmatrix}0 & 0 & 0\\ 0 & 0 & i\gamma\\ 0 & -i\gamma & 0\end{pmatrix} - \omega^2\zeta^2\begin{pmatrix}1 & 0 & 0\\ 0 & 0 & 0\\ 0 & 0 & 0\end{pmatrix}$ | $(\nabla\times)\vert_{\partial_x=0}$ | $\dfrac{1}{i\omega\mu}(\nabla\times) + \dfrac{\zeta}{\mu}\begin{pmatrix}1 & 0 & 0\\ 0 & 0 & 0\\ 0 & 0 & 0\end{pmatrix}$ |
| G-c | $-\omega^2\mu\varepsilon\,\mathbf{I} - \omega^2\mu\begin{pmatrix}0 & 0 & -i\gamma\\ 0 & 0 & 0\\ i\gamma & 0 & 0\end{pmatrix} - \omega^2\zeta^2\begin{pmatrix}1 & 0 & 0\\ 0 & 0 & 0\\ 0 & 0 & 0\end{pmatrix}$ | $(\nabla\times)\vert_{\partial_x=0}$ | $\dfrac{1}{i\omega\mu}(\nabla\times) + \dfrac{\zeta}{\mu}\begin{pmatrix}1 & 0 & 0\\ 0 & 0 & 0\\ 0 & 0 & 0\end{pmatrix}$ |
| H-a | $-\omega^2\mu\varepsilon\,\mathbf{I} - \omega^2\mu\begin{pmatrix}0 & i\gamma & 0\\ -i\gamma & 0 & 0\\ 0 & 0 & 0\end{pmatrix} - \omega^2\zeta^2\begin{pmatrix}0 & 0 & 0\\ 0 & 1 & 0\\ 0 & 0 & 0\end{pmatrix}$ | $(\nabla\times)\vert_{\partial_y=0}$ | $\dfrac{1}{i\omega\mu}(\nabla\times) + \dfrac{\zeta}{\mu}\begin{pmatrix}0 & 0 & 0\\ 0 & 1 & 0\\ 0 & 0 & 0\end{pmatrix}$ |

| | | | | |
|---|---|---|---|---|
| H-b | $-\omega^2\mu\varepsilon\,\mathbf{I} - \omega^2\mu\begin{pmatrix}0 & 0 & 0\\ 0 & 0 & i\gamma\\ 0 & -i\gamma & 0\end{pmatrix} - \omega^2\zeta^2\begin{pmatrix}0 & 0 & 0\\ 0 & 1 & 0\\ 0 & 0 & 0\end{pmatrix}$ | $(\nabla\times)\vert_{\partial_y=0}$ | $\frac{1}{i\omega\mu}(\nabla\times) + \frac{\zeta}{\mu}\begin{pmatrix}0 & 0 & 0\\ 0 & 1 & 0\\ 0 & 0 & 0\end{pmatrix}$ | |
| H-c | $-\omega^2\mu\varepsilon\,\mathbf{I} - \omega^2\mu\begin{pmatrix}0 & 0 & -i\gamma\\ 0 & 0 & 0\\ i\gamma & 0 & 0\end{pmatrix} - \omega^2\zeta^2\begin{pmatrix}0 & 0 & 0\\ 0 & 1 & 0\\ 0 & 0 & 0\end{pmatrix}$ | $(\nabla\times)\vert_{\partial_y=0}$ | $\frac{1}{i\omega\mu}(\nabla\times) + \frac{\zeta}{\mu}\begin{pmatrix}0 & 0 & 0\\ 0 & 1 & 0\\ 0 & 0 & 0\end{pmatrix}$ | |
| I-a | $-\omega^2\mu\varepsilon\,\mathbf{I} - \omega^2\mu\begin{pmatrix}0 & i\gamma & 0\\ -i\gamma & 0 & 0\\ 0 & 0 & 0\end{pmatrix} - \omega^2\zeta^2\begin{pmatrix}0 & 0 & 0\\ 0 & 0 & 0\\ 0 & 0 & 1\end{pmatrix}$ | $(\nabla\times)\vert_{\partial_z=0}$ | $\frac{1}{i\omega\mu}(\nabla\times) + \frac{\zeta}{\mu}\begin{pmatrix}0 & 0 & 0\\ 0 & 0 & 0\\ 0 & 0 & 1\end{pmatrix}$ | |
| I-b | $-\omega^2\mu\varepsilon\,\mathbf{I} - \omega^2\mu\begin{pmatrix}0 & 0 & 0\\ 0 & 0 & i\gamma\\ 0 & -i\gamma & 0\end{pmatrix} - \omega^2\zeta^2\begin{pmatrix}0 & 0 & 0\\ 0 & 0 & 0\\ 0 & 0 & 1\end{pmatrix}$ | $(\nabla\times)\vert_{\partial_z=0}$ | $\frac{1}{i\omega\mu}(\nabla\times) + \frac{\zeta}{\mu}\begin{pmatrix}0 & 0 & 0\\ 0 & 0 & 0\\ 0 & 0 & 1\end{pmatrix}$ | |
| I-c | $-\omega^2\mu\varepsilon\,\mathbf{I} - \omega^2\mu\begin{pmatrix}0 & 0 & -i\gamma\\ 0 & 0 & 0\\ i\gamma & 0 & 0\end{pmatrix} - \omega^2\zeta^2\begin{pmatrix}0 & 0 & 0\\ 0 & 0 & 0\\ 0 & 0 & 1\end{pmatrix}$ | $(\nabla\times)\vert_{\partial_z=0}$ | $\frac{1}{i\omega\mu}(\nabla\times) + \frac{\zeta}{\mu}\begin{pmatrix}0 & 0 & 0\\ 0 & 0 & 0\\ 0 & 0 & 1\end{pmatrix}$ | |

## 4. Study cases

In this section, the extended wave equation obtained in the previous section is applied to several free space and waveguide models. In particular, the following models will be discussed in each of the sections.

4.1 Waveguide propagation without both MO and ME effects
4.2 Free space propagation with only MO effect
4.3 Waveguide propagation with only MO effect
4.4 Free space propagation with only ME effect
4.5 Waveguide propagation with only ME effect
4.6 Free space propagation with both MO and ME effects
4.7 Waveguide propagation with both MO and ME effects

### 4.1 Waveguide propagation without both MO and ME effects

The case of light propagating in a waveguide where neither MO nor ME effects are first considered. For simplicity, a slab waveguide ($\partial_x = 0, \partial_z = -j\beta$) in the TM mode ($H_y = 0, H_z = 0, E_x = 0$) is assumed, and Eqs. (11a) and (11b) can be written as follows.

$$\left((\nabla\times\nabla\times) + \mathbf{M}_1\right)\begin{pmatrix}H_x\\ 0\\ 0\end{pmatrix} = 0 \tag{15a}$$

$$\begin{pmatrix}0\\ E_y\\ E_z\end{pmatrix} = \mathbf{Z}_1^{-1}\mathbf{Z}_2\begin{pmatrix}H_x\\ 0\\ 0\end{pmatrix} \tag{15b}$$

Since we can express the equation as

$$(\nabla\times\nabla\times) = \begin{pmatrix}-\partial_y\partial_y-\partial_z\partial_z & \partial_x\partial_y & \partial_x\partial_z\\ \partial_x\partial_y & -\partial_x\partial_x-\partial_z\partial_z & \partial_y\partial_z\\ \partial_x\partial_z & \partial_y\partial_z & -\partial_x\partial_x-\partial_y\partial_y\end{pmatrix} = \begin{pmatrix}-\partial_y\partial_y+\beta^2 & 0 & 0\\ 0 & \beta^2 & -i\beta\partial_y\\ 0 & -i\beta\partial_y & -\partial_y\partial_y\end{pmatrix} \tag{16a}$$

$$\mathbf{M}_1 = -\omega^2\varepsilon\mu\,\mathbf{I} \tag{16b}$$

$$\mathbf{Z}_1 = \mathbf{I} \tag{16c}$$

$$Z_2 = -\frac{1}{i\omega\varepsilon}(\nabla\times) = -\frac{1}{i\omega\varepsilon}\begin{pmatrix} 0 & -\partial_z & \partial_y \\ \partial_z & 0 & -\partial_x \\ -\partial_y & \partial_x & 0 \end{pmatrix} = -\frac{1}{i\omega\varepsilon}\begin{pmatrix} 0 & i\beta & \partial_y \\ -i\beta & 0 & 0 \\ -\partial_y & 0 & 0 \end{pmatrix} \quad (16d)$$

$$\partial_y\partial_y H_x + (\omega^2\varepsilon\mu - \beta^2)H_x = 0 \quad (16a)$$

$$E_z = \frac{1}{i\omega\varepsilon}\partial_y H_x \quad (16b)$$

Using a similar procedure and assuming a slab waveguide ($\partial_x = 0, \partial_z = -i\beta$) in the TE mode ($E_y = 0, E_z = 0, H_x = 0$), we can verify that Eqs. (13a) and (13b) can be attributed to the following general equations for optical waveguide.

$$\partial_y\partial_y E_x + (\omega^2\varepsilon\mu - \beta^2)E_x = 0 \quad (17a)$$

$$H_z = -\frac{1}{i\omega\mu}\partial_y E_x \quad (17b)$$

### 4.2 Free space propagation with only MO effect

It is assumed that plane waves propagate in a uniform crystal ($\partial_x = 0, \partial_y = 0, \partial_z = -i\beta$) and only MO effect is considered. Eqs. (13a) and (13b) are given as follows.

$$\big((\nabla\times\nabla\times) + \mathbf{N}_1\big)\begin{pmatrix} E_x \\ E_y \\ 0 \end{pmatrix} = 0 \quad (18a)$$

$$\begin{pmatrix} H_x \\ H_y \\ 0 \end{pmatrix} = \mathbf{Y}\begin{pmatrix} E_x \\ E_y \\ 0 \end{pmatrix} \quad (18b)$$

Assuming that the magnetization is aligned in the z direction ($a = i\gamma, b = 0, c = 0$), we can write

$$(\nabla\times\nabla\times) = \begin{pmatrix} -\partial_y\partial_y-\partial_z\partial_z & \partial_x\partial_y & \partial_x\partial_z \\ \partial_x\partial_y & -\partial_x\partial_x-\partial_z\partial_z & \partial_y\partial_z \\ \partial_x\partial_z & \partial_y\partial_z & -\partial_x\partial_x-\partial_y\partial_y \end{pmatrix} = \begin{pmatrix} \beta^2 & 0 & 0 \\ 0 & \beta^2 & 0 \\ 0 & 0 & 0 \end{pmatrix} \quad (19a)$$

$$\mathbf{N}_1 = -\omega^2\varepsilon\mu\mathbf{I} - \omega^2\mu\begin{pmatrix} 0 & i\gamma & 0 \\ -i\gamma & 0 & 0 \\ 0 & 0 & 0 \end{pmatrix} \quad (19b)$$

for each of the cases (A~I-a) in Table 2. We substitute these into Eqs. (18a) and (18b), and obtain

$$\begin{pmatrix} \beta^2 - \omega^2\varepsilon\mu & -i\omega^2\mu\gamma \\ i\omega^2\mu\gamma & \beta^2 - \omega^2\varepsilon\mu \end{pmatrix}\begin{pmatrix} E_x \\ E_y \end{pmatrix} = 0 \quad (20)$$

Finally, by diagonalizing Eq. (20), we obtain the following equation.

$$\begin{pmatrix} \beta^2 - \omega^2\varepsilon\mu + \omega^2\mu\gamma & 0 \\ 0 & \beta^2 - \omega^2\varepsilon\mu - \omega^2\mu\gamma \end{pmatrix}\begin{pmatrix} \frac{i}{2}(E_x - iE_y) \\ \frac{i}{2}(-E_x - iE_y) \end{pmatrix} = 0 \quad (21)$$

where $E_x - iE_y$ represents RCP, and $-E_x - iE_y$ represents LCP. Thus, the fundamental modes of propagating light are LCP and RCP modes. The propagation constants for each mode are given by

$$\beta_\pm = \omega\sqrt{\mu(\varepsilon \pm \gamma)} \quad (22)$$

This leads to a nonreciprocal polarization rotation of the linearly polarized light in the crystal, which is consistent with the theory of the Faraday effect [40,41].

### 4.3 Waveguide propagation with only MO effect

Under the assumption of a slab waveguide ($\partial_x = 0, \partial_z = -i\beta$) in the TM mode ($H_y = 0, H_z = 0, E_x = 0$), only MO effect is considered. Eqs. (11a) and (11b) are expressed as follows.

$$\left((\nabla \times \nabla \times) + \mathbf{M}_1 + \frac{\gamma^2}{\epsilon^2}\mathbf{M}_3 + \frac{i\gamma}{\epsilon}\mathbf{M}_4\right)\begin{pmatrix} H_x \\ 0 \\ 0 \end{pmatrix} = 0 \quad (23a)$$

$$\begin{pmatrix} 0 \\ E_y \\ E_z \end{pmatrix} = \mathbf{Z}_1^{-1}\mathbf{Z}_2 \begin{pmatrix} H_x \\ 0 \\ 0 \end{pmatrix} \quad (23b)$$

Assuming that the magnetization is aligned in the $x$ direction ($a = 0, b = i\gamma, c = 0$), we can write

$$(\nabla \times \nabla \times) = \begin{pmatrix} -\partial_y\partial_y - \partial_z\partial_z & \partial_x\partial_y & \partial_x\partial_z \\ \partial_x\partial_y & -\partial_x\partial_x - \partial_z\partial_z & \partial_y\partial_z \\ \partial_x\partial_z & \partial_y\partial_z & -\partial_x\partial_x - \partial_y\partial_y \end{pmatrix} = \begin{pmatrix} -\partial_y\partial_y + \beta^2 & 0 & 0 \\ 0 & \beta^2 & -i\beta\partial_y \\ 0 & -i\beta\partial_y & -\partial_y\partial_y \end{pmatrix} \quad (24a)$$

$$\mathbf{M}_1 = -\omega^2\varepsilon\mu\left(1 - \frac{\gamma^2}{\epsilon^2}\right)\mathbf{I} \quad (24b)$$

$$\mathbf{M}_3 = \begin{pmatrix} 0 & 0 & 0 \\ 0 & \partial_z\partial_z & -\partial_y\partial_z \\ 0 & -\partial_y\partial_z & \partial_y\partial_y \end{pmatrix} = \begin{pmatrix} 0 & 0 & 0 \\ 0 & -\beta^2 & i\beta\partial_y \\ 0 & i\beta\partial_y & \partial_y\partial_y \end{pmatrix} \quad (24c)$$

$$\mathbf{M}_4 = \partial_x(\nabla \times) = 0 \quad (24d)$$

$$\mathbf{Z}_1 = \begin{pmatrix} 1 & 0 & 0 \\ 0 & 1 & \frac{i\gamma}{\epsilon} \\ 0 & -\frac{i\gamma}{\epsilon} & 1 \end{pmatrix} \quad (24e)$$

$$\mathbf{Z}_2 = -\frac{1}{i\omega\varepsilon}(\nabla\times) = -\frac{1}{i\omega\varepsilon}\begin{pmatrix} 0 & -\partial_z & \partial_y \\ \partial_z & 0 & -\partial_x \\ -\partial_y & \partial_x & 0 \end{pmatrix} = -\frac{1}{i\omega\varepsilon}\begin{pmatrix} 0 & i\beta & \partial_y \\ -i\beta & 0 & 0 \\ -\partial_y & 0 & 0 \end{pmatrix} \quad (24f)$$

for each of the cases (A~I-b) in Table 1. These are substituted into Eqs. (23a) and (23b), and give

$$\partial_y\partial_y H_x + \left(\omega^2\varepsilon\mu\left(1 - \frac{\gamma^2}{\epsilon^2}\right) - \beta^2\right)H_x = 0 \quad (25a)$$

$$E_z = \frac{1}{i\omega}\frac{\varepsilon}{\epsilon^2 - \gamma^2}\left(\partial_y H_x + \frac{\gamma\beta}{\epsilon}H_x\right) \quad (25b)$$

These equations are consistent with the wave equation for a waveguide optical isolator using the transverse magneto-optical Kerr effect [42–46]. Equation (25b) shows that the first-order term of the parameter $\gamma$ for MO effect is multiplied by the first-order term of the propagation constant $\beta$. This gives the nonreciprocal propagation constant (nonreciprocal phase shifts (NRPS), nonreciprocal loss (NRL), etc.) in the forward and backward waves.

### 4.4 Free space propagation with only ME effect

Assume that plane waves propagate in a uniform crystal ($\partial_x = 0, \partial_y = 0, \partial_z = -i\beta$) and only ME effect is considered. Equations. (13a) and (13b) are expressed as follows.

$$\left((\nabla \times \nabla \times) + \mathbf{N}_1 + i\omega\zeta\mathbf{N}_2\right)\begin{pmatrix} E_x \\ E_y \\ 0 \end{pmatrix} = 0 \quad (26a)$$

$$\begin{pmatrix} H_x \\ H_y \\ 0 \end{pmatrix} = \mathbf{Y}\begin{pmatrix} E_x \\ E_y \\ 0 \end{pmatrix} \quad (26b)$$

Assuming that the diagonal components $G$ and $H$ contain values ($G = H = \zeta$), Table 2 leads to

$$(\nabla \times \nabla \times) = \begin{pmatrix} -\partial_y\partial_y - \partial_z\partial_z & \partial_x\partial_y & \partial_x\partial_z \\ \partial_x\partial_y & -\partial_x\partial_x - \partial_z\partial_z & \partial_y\partial_z \\ \partial_x\partial_z & \partial_y\partial_z & -\partial_x\partial_x - \partial_y\partial_y \end{pmatrix} = \begin{pmatrix} \beta^2 & 0 & 0 \\ 0 & \beta^2 & 0 \\ 0 & 0 & 0 \end{pmatrix} \quad (27a)$$

$$\mathbf{N}_1 = -\omega^2 \varepsilon\mu \mathbf{I} - \omega^2 \zeta^2 \begin{pmatrix} 1 & 0 & 0 \\ 0 & 1 & 0 \\ 0 & 0 & 0 \end{pmatrix} \quad (27b)$$

$$\mathbf{N}_2 = (\nabla \times)|_{\partial_x = 0} = \begin{pmatrix} 0 & -\partial_z & \partial_y \\ \partial_z & 0 & 0 \\ -\partial_y & 0 & 0 \end{pmatrix} = \begin{pmatrix} 0 & i\beta & 0 \\ -i\beta & 0 & 0 \\ 0 & 0 & 0 \end{pmatrix} \quad (27c)$$

Substituting these into Eqs. (26a) and (26b), we obtain

$$\begin{pmatrix} \beta^2 - \omega^2\varepsilon\mu - \omega^2\zeta^2 & -\omega\beta\zeta \\ \omega\beta\zeta & \beta^2 - \omega^2\varepsilon\mu - \omega^2\zeta^2 \end{pmatrix} \begin{pmatrix} E_x \\ E_y \end{pmatrix} = 0 \quad (28)$$

Finally, by diagonalizing Eq. (28), the following equation is obtained.

$$\begin{pmatrix} \beta^2 - \omega^2\varepsilon\mu - \omega^2\zeta^2 - i\omega\beta\zeta & 0 \\ 0 & \beta^2 - \omega^2\varepsilon\mu - \omega^2\zeta^2 + i\omega\beta\zeta \end{pmatrix} \begin{pmatrix} \frac{i}{2}(E_x - iE_y) \\ \frac{i}{2}(-E_x - iE_y) \end{pmatrix} = 0 \quad (29)$$

where $E_x - iE_y$ represents RCP, and $-E_x - iE_y$ represents LCP. Thus, the fundamental modes of propagating light are LCP and RCP modes. The propagation constants for each mode are given by

$$\beta_\pm = \sqrt{\omega^2\varepsilon\mu + \omega^2\zeta^2 \pm i\omega\beta\zeta} \quad (30)$$

This result is consistent with the equations for plane wave propagation in bulk crystals with ME effect derived by Engheta et al. in 1992 [30,31] and Ioannidis et al. in 2010 [47].

### 4.5 Waveguide propagation with only ME effect

Assume a slab waveguide ($\partial_x = 0, \partial_z = -i\beta$) in the TM mode ($H_y = 0, H_z = 0, E_x = 0$) and only ME effect is considered. Equations. (11a) and (11b) are expressed as follows.

$$((\nabla \times \nabla \times) + \mathbf{M}_1 + i\omega\zeta\mathbf{M}_2) \begin{pmatrix} H_x \\ 0 \\ 0 \end{pmatrix} = 0 \quad (31a)$$

$$\begin{pmatrix} 0 \\ E_y \\ E_z \end{pmatrix} = \mathbf{Z}_1^{-1}\mathbf{Z}_2 \begin{pmatrix} H_x \\ 0 \\ 0 \end{pmatrix} \quad (31b)$$

Assuming that the diagonal component $C$ contains value ($C = \zeta$), we can write

$$(\nabla \times \nabla \times) = \begin{pmatrix} -\partial_y\partial_y - \partial_z\partial_z & \partial_x\partial_y & \partial_x\partial_z \\ \partial_x\partial_y & -\partial_x\partial_x - \partial_z\partial_z & \partial_y\partial_z \\ \partial_x\partial_z & \partial_y\partial_z & -\partial_x\partial_x - \partial_y\partial_y \end{pmatrix} = \begin{pmatrix} -\partial_y\partial_y + \beta^2 & 0 & 0 \\ 0 & \beta^2 & -i\beta\partial_y \\ 0 & -i\beta\partial_y & -\partial_y\partial_y \end{pmatrix} \quad (32a)$$

$$\mathbf{M}_1 = -\omega^2\varepsilon\mu\mathbf{I} - \omega^2\zeta^2 \begin{pmatrix} 1 & 0 & 0 \\ 0 & 0 & 0 \\ 0 & 0 & 0 \end{pmatrix} \quad (32b)$$

$$\mathbf{M}_2 = \begin{pmatrix} 0 & \partial_x & 0 \\ -\partial_x & 0 & 0 \\ 0 & 0 & 0 \end{pmatrix} = \begin{pmatrix} 0 & 0 & 0 \\ 0 & 0 & 0 \\ 0 & 0 & 0 \end{pmatrix} \quad (32c)$$

$$\mathbf{Z}_1 = \begin{pmatrix} 1 & 0 & 0 \\ 0 & 1 & 0 \\ 0 & 0 & 1 \end{pmatrix} \quad (32d)$$

$$Z_2 = -\frac{1}{i\omega\varepsilon}(\nabla\times) - \frac{\zeta}{\varepsilon}\begin{pmatrix}0&0&0\\0&0&0\\1&0&0\end{pmatrix} = -\frac{1}{i\omega\varepsilon}\begin{pmatrix}0&-\partial_z&\partial_y\\\partial_z&0&-\partial_x\\-\partial_y&\partial_x&0\end{pmatrix} - \frac{\zeta}{\varepsilon}\begin{pmatrix}0&0&0\\0&0&0\\1&0&0\end{pmatrix}$$

$$= -\frac{1}{i\omega\varepsilon}\begin{pmatrix}0&i\beta&\partial_y\\-i\beta&0&0\\-\partial_y&0&0\end{pmatrix} - \frac{\zeta}{\varepsilon}\begin{pmatrix}0&0&0\\0&0&0\\1&0&0\end{pmatrix} = -\frac{1}{i\omega\varepsilon}\begin{pmatrix}0&i\beta&\partial_y\\-i\beta&0&0\\-\partial_y+i\omega\zeta&0&0\end{pmatrix} \quad (32e)$$

for each of the cases (C-a~c) in Table 1. These are substituted into Eqs. (31a) and (31b), and give

$$\partial_y\partial_y H_x + (\omega^2\varepsilon\mu + \omega^2\zeta^2 - \beta^2)H_x = 0 \tag{33a}$$

$$E_z = \frac{1}{i\omega\varepsilon}(\partial_y H_x - i\omega\zeta H_x) \tag{33b}$$

In the above case, we can keep the TM mode in the waveguide, but not depending on how each of the off-diagonal components $A$ - $F$ of the magnetic-to-electric coupling enters. For example, when $A$ and $D$ have values, the three components of the TE mode ($E_x, H_y, H_z$) interact with each other, to maintain the TE mode in the waveguide. When $C$ and $E$ have values, the three components of the TM mode ($H_x, E_y, E_z$) interact with each other, to maintain the TM mode in the waveguide. In contrast, when $B$ and $F$ have values, the three components of the TE mode and the three components of the TM mode interfere with each other. In this case, a special situation arises where both TM and TE modes are mixed in the waveguide.

### 4.6 Free space propagation with both MO and ME effects

Assume that plane waves propagate in a uniform crystal ($\partial_x = 0, \partial_y = 0, \partial_z = -i\beta$) and consider both MO and ME effects are considered. Eqs. (13a) and (13b) are expressed as follows.

$$((\nabla\times\nabla\times) + N_1 + i\omega\zeta N_2)\begin{pmatrix}E_x\\E_y\\0\end{pmatrix} = 0 \tag{34a}$$

$$\begin{pmatrix}H_x\\H_y\\0\end{pmatrix} = Y\begin{pmatrix}E_x\\E_y\\0\end{pmatrix} \tag{34b}$$

Assuming that the diagonal components $G$ and $H$ contain values ($G = H = \zeta$) and the magnetization is aligned in the $z$ direction ($a = i\gamma, b = 0, c = 0$), Table 2 leads to

$$(\nabla\times\nabla\times) = \begin{pmatrix}-\partial_y\partial_y-\partial_z\partial_z & \partial_x\partial_y & \partial_x\partial_z\\ \partial_x\partial_y & -\partial_x\partial_x-\partial_z\partial_z & \partial_y\partial_z\\ \partial_x\partial_z & \partial_y\partial_z & -\partial_x\partial_x-\partial_y\partial_y\end{pmatrix} = \begin{pmatrix}\beta^2 & 0 & 0\\ 0 & \beta^2 & 0\\ 0 & 0 & 0\end{pmatrix} \tag{35a}$$

$$N_1 = -\omega^2\varepsilon\mu I - \omega^2\mu\begin{pmatrix}0 & i\gamma & 0\\ -i\gamma & 0 & 0\\ 0 & 0 & 0\end{pmatrix} - \omega^2\zeta^2\begin{pmatrix}1&0&0\\0&1&0\\0&0&0\end{pmatrix} \tag{35b}$$

$$N_2 = (\nabla\times)|_{\partial_x=0} = \begin{pmatrix}0&-\partial_z&\partial_y\\\partial_z&0&0\\-\partial_y&0&0\end{pmatrix} = \begin{pmatrix}0&i\beta&0\\-i\beta&0&0\\0&0&0\end{pmatrix} \tag{35c}$$

Substituting these into Eqs. (34a) and (34b), we obtain

$$\begin{pmatrix}\beta^2-\omega^2\varepsilon\mu-\omega^2\zeta^2 & -\omega\beta\zeta-i\omega^2\mu\gamma\\ \omega\beta\zeta+i\omega^2\mu\gamma & \beta^2-\omega^2\varepsilon\mu-\omega^2\zeta^2\end{pmatrix}\begin{pmatrix}E_x\\E_y\end{pmatrix} = 0 \tag{36}$$

Finally, by diagonalizing Eq. (36), the following equation is obtained.

$$\begin{pmatrix}\beta^2-\omega^2\varepsilon\mu-\omega^2\zeta^2-i\omega\beta\zeta+\omega^2\mu\gamma & 0\\ 0 & \beta^2-\omega^2\varepsilon\mu-\omega^2\zeta^2+i\omega\beta\zeta-\omega^2\mu\gamma\end{pmatrix}\begin{pmatrix}\frac{i}{2}(E_x-iE_y)\\ \frac{i}{2}(-E_x-iE_y)\end{pmatrix} = 0 \tag{37}$$

where $E_x - iE_y$ represents RCP, and $-E_x - iE_y$ represents LCP. Thus, the fundamental modes of the propagating light are LCP and RCP modes. The propagation constants for each mode are obtained by solving the following quadratic equation.

$$f_+(\beta) = \beta^2 - i\omega\beta\zeta - \omega^2\varepsilon\mu - \omega^2\zeta^2 + \omega^2\mu\gamma = 0 \tag{38a}$$
$$f_-(\beta) = \beta^2 + i\omega\beta\zeta - \omega^2\varepsilon\mu - \omega^2\zeta^2 - \omega^2\mu\gamma = 0 \tag{38b}$$

Equations (38) shows that the first-order terms of the parameter $\gamma$ for MO effect and the parameter $\zeta$ for ME effect are multiplied by the first-order term of the propagation constant $\beta$. This implies that nonreciprocal polarization rotation (as opposed to nonreciprocal polarization rotation, as described in Section 4.2) is possible in the forward and backward waves.

### 4.7 Waveguide propagation with both MO and ME effects

Assume a slab waveguide ($\partial_x = 0, \partial_z = -i\beta$) in the TM mode ($H_y = 0, H_z = 0, E_x = 0$) and consider both MO and ME effects. Equations. (11a) and (11b) are expressed as follows.

$$\left((\nabla \times \nabla \times) + \mathbf{M}_1 + i\omega\zeta\mathbf{M}_2 + \frac{\gamma^2}{\epsilon^2}\mathbf{M}_3 + \frac{i\gamma}{\epsilon}\mathbf{M}_4 + \frac{\omega\zeta\gamma}{\epsilon}\mathbf{M}_5\right)\begin{pmatrix}H_x\\0\\0\end{pmatrix} = 0 \tag{39a}$$

$$\begin{pmatrix}0\\E_y\\E_z\end{pmatrix} = \mathbf{Z}_1^{-1}\mathbf{Z}_2\begin{pmatrix}H_x\\0\\0\end{pmatrix} \tag{39b}$$

Assuming that the diagonal component $C$ contains value ($C = \zeta$) and the magnetization is aligned in the $x$ direction ($a = 0, b = i\gamma, c = 0$), Table 1 (C-b) leads to

$$(\nabla\times\nabla\times) = \begin{pmatrix}-\partial_y\partial_y-\partial_z\partial_z & \partial_x\partial_y & \partial_x\partial_z\\ \partial_x\partial_y & -\partial_x\partial_x-\partial_z\partial_z & \partial_y\partial_z\\ \partial_x\partial_z & \partial_y\partial_z & -\partial_x\partial_x-\partial_y\partial_y\end{pmatrix} = \begin{pmatrix}-\partial_y\partial_y+\beta^2 & 0 & 0\\ 0 & \beta^2 & -i\beta\partial_y\\ 0 & -i\beta\partial_y & -\partial_y\partial_y\end{pmatrix} \tag{40a}$$

$$\mathbf{M}_1 = -\omega^2\varepsilon\mu\left(1-\frac{\gamma^2}{\epsilon^2}\right)\mathbf{I} - \omega^2\zeta^2\begin{pmatrix}1 & 0 & 0\\0 & 0 & 0\\0 & 0 & 0\end{pmatrix} \tag{40b}$$

$$\mathbf{M}_2 = \begin{pmatrix}0 & \partial_x & 0\\-\partial_x & 0 & 0\\0 & 0 & 0\end{pmatrix} = \begin{pmatrix}0 & 0 & 0\\0 & 0 & 0\\0 & 0 & 0\end{pmatrix} \tag{40c}$$

$$\mathbf{M}_3 = \begin{pmatrix}0 & 0 & 0\\0 & \partial_z\partial_z & -\partial_y\partial_z\\0 & -\partial_y\partial_z & \partial_y\partial_y\end{pmatrix} = \begin{pmatrix}0 & 0 & 0\\0 & -\beta^2 & i\beta\partial_y\\0 & i\beta\partial_y & \partial_y\partial_y\end{pmatrix} \tag{40d}$$

$$\mathbf{M}_4 = -\partial_x(\nabla\times) = 0 \tag{40e}$$

$$\mathbf{M}_5 = \begin{pmatrix}-2\partial_z & 0 & \partial_x\\0 & 0 & 0\\\partial_x & 0 & 0\end{pmatrix} = \begin{pmatrix}2i\beta & 0 & 0\\0 & 0 & 0\\0 & 0 & 0\end{pmatrix} \tag{40f}$$

$$\mathbf{Z}_1 = \begin{pmatrix}1 & 0 & 0\\0 & 1 & \frac{i\gamma}{\epsilon}\\0 & -\frac{i\gamma}{\epsilon} & 1\end{pmatrix} \tag{40g}$$

$$\mathbf{Z}_2 = -\frac{1}{i\omega\varepsilon}(\nabla\times) - \frac{\zeta}{\varepsilon}\begin{pmatrix}0 & 0 & 0\\0 & 0 & 0\\1 & 0 & 0\end{pmatrix} = -\frac{1}{i\omega\varepsilon}\begin{pmatrix}0 & -\partial_z & \partial_y\\ \partial_z & 0 & -\partial_x\\ -\partial_y & \partial_x & 0\end{pmatrix} - \frac{\zeta}{\varepsilon}\begin{pmatrix}0 & 0 & 0\\0 & 0 & 0\\1 & 0 & 0\end{pmatrix}$$

$$= -\frac{1}{i\omega\varepsilon}\begin{pmatrix}0 & i\beta & \partial_y\\-i\beta & 0 & 0\\-\partial_y & 0 & 0\end{pmatrix} - \frac{\zeta}{\varepsilon}\begin{pmatrix}0 & 0 & 0\\0 & 0 & 0\\1 & 0 & 0\end{pmatrix} = -\frac{1}{i\omega\varepsilon}\begin{pmatrix}0 & i\beta & \partial_y\\-i\beta & 0 & 0\\-\partial_y + i\omega\zeta & 0 & 0\end{pmatrix} \tag{40h}$$

Substituting these into Eqs. (39a) and (39b), we obtain

$$\partial_y\partial_y H_x + \left(\omega^2\varepsilon\mu\left(1-\frac{\gamma^2}{\epsilon^2}\right) + \omega^2\zeta^2 - \beta^2 - 2i\beta\frac{\omega\zeta\gamma}{\epsilon}\right)H_x = 0 \qquad (41a)$$

$$E_z = \frac{1}{i\omega}\frac{\varepsilon}{\epsilon^2-\gamma^2}\left(\frac{\gamma\beta}{\varepsilon}H_x + \partial_y H_x - i\omega\zeta H_x\right) \qquad (41b)$$

Equations (41) shows that the first-order term of the parameter $\gamma$ for MO effect is multiplied by the first-order term of the propagation constant $\beta$. It also shows that the parameter $\zeta$ related to ME effect functions in such a way to increase or decrease that effect. This allows for the amplification of nonreciprocal effects (e.g., NRPS, NRS, etc.) within waveguides.

Note that the interaction term $\zeta\gamma$ of MO and ME effects, as in Eqs. (41), occurs only in certain cases. For example, if we consider a waveguide with metamaterial structure (Fig. 2) and a ferromagnetic medium (Fig. 1) in which MO and ME effects can be controlled independently, the interaction terms of MO and ME effects appear when the magnetization direction and the direction of the magnetic moment of the SRR coincide (b-C, b-E).

In waveguide propagation considering both MO and ME effects, as described in Section 4.5, it may be difficult to maintain TM and TE modes in the waveguide depending on the value of each element of the off-diagonal component $A$-$F$ of the magneto-electric coupling tensor. Specifically, when $A$ and $D$ have values, the TE mode is maintained, and when $C$ and $E$ have values, the TM mode is maintained. In contrast, when $B$ and $F$ have values, both TE and TM modes interfere with each other in the waveguide in a complicated manner.

Finally, Table 3 summarizes the wave equations derived in Sections 4.1 through 4.7 for each case.

**Table 3. Wave equations for MO effect and ME effect**

| Related effects | Propagation | MO tensor | ME tensor | Wave equation | Section |
|---|---|---|---|---|---|
| **General** | Waveguide | 0 | 0 | $\partial_y\partial_y H_x + (\omega^2\varepsilon\mu - \beta^2)H_x = 0$ <br> $E_z = \frac{1}{i\omega\varepsilon}\partial_y H_x$ | 4.1 |
| **Faraday effect** | Free space | $\begin{pmatrix}0 & i\gamma & 0\\-i\gamma & 0 & 0\\0 & 0 & 0\end{pmatrix}$ | 0 | $\begin{pmatrix}\beta^2 - \omega^2\varepsilon\mu + \omega^2\mu\gamma & 0 \\ 0 & \beta^2 - \omega^2\varepsilon\mu - \omega^2\mu\gamma\end{pmatrix}\begin{pmatrix}\frac{i}{2}(E_x - iE_y)\\\frac{i}{2}(-E_x - iE_y)\end{pmatrix} = 0$ | 4.2 |
| **Transverse Kerr effect** | Waveguide (TM mode) | $\begin{pmatrix}0 & 0 & -i\gamma\\0 & 0 & 0\\i\gamma & 0 & 0\end{pmatrix}$ | 0 | $\partial_y\partial_y H_x + \left(\omega^2\varepsilon\mu\left(1-\frac{\gamma^2}{\epsilon^2}\right) - \beta^2\right)H_x = 0$ <br> $E_z = \frac{1}{i\omega}\frac{\varepsilon}{\epsilon^2-\gamma^2}\left(\partial_y H_x + \frac{\gamma\beta}{\epsilon}H_x\right)$ | 4.3 |
| **ME effect** | Free space | 0 | $\begin{pmatrix}\zeta & 0 & 0\\0 & \zeta & 0\\0 & 0 & 0\end{pmatrix}$ | $\begin{pmatrix}\beta^2 - \omega^2\varepsilon\mu - \omega^2\zeta^2 - i\omega\beta\zeta & 0 \\ 0 & \beta^2 - \omega^2\varepsilon\mu - \omega^2\zeta^2 + i\omega\beta\zeta\end{pmatrix}\begin{pmatrix}\frac{i}{2}(E_x - iE_y)\\\frac{i}{2}(-E_x - iE_y)\end{pmatrix} = 0$ | 4.4 |
| **ME effect** | Waveguide (TM mode) | 0 | $\begin{pmatrix}0 & 0 & 0\\0 & 0 & \zeta\\0 & 0 & 0\end{pmatrix}$ | $\partial_y\partial_y H_x + (\omega^2\varepsilon\mu + \omega^2\zeta^2 - \beta^2)H_x = 0$ <br> $E_z = \frac{1}{i\omega\varepsilon}(\partial_y H_x - i\omega\zeta H_x)$ | 4.5 |
| **MO&ME effect** | Free space | $\begin{pmatrix}0 & i\gamma & 0\\-i\gamma & 0 & 0\\0 & 0 & 0\end{pmatrix}$ | $\begin{pmatrix}\zeta & 0 & 0\\0 & \zeta & 0\\0 & 0 & 0\end{pmatrix}$ | $\begin{pmatrix}\beta^2 - \omega^2\varepsilon\mu - \omega^2\zeta^2 - i\omega\beta\zeta + \omega^2\mu\gamma & 0 \\ 0 & \beta^2 - \omega^2\varepsilon\mu - \omega^2\zeta^2 + i\omega\beta\zeta - \omega^2\mu\gamma\end{pmatrix}\begin{pmatrix}\frac{i}{2}(E_x - iE_y)\\\frac{i}{2}(-E_x - iE_y)\end{pmatrix}$ <br> $= 0$ | 4.6 |
| **MO&ME effects** | Waveguide (TM mode) | $\begin{pmatrix}0 & i\gamma & 0\\-i\gamma & 0 & 0\\0 & 0 & 0\end{pmatrix}$ | $\begin{pmatrix}0 & 0 & 0\\0 & 0 & 0\\\zeta & 0 & 0\end{pmatrix}$ | $\partial_y\left(1-\frac{\gamma^2}{\epsilon^2}\right)H_x + \left(\omega^2\left(1-\frac{\gamma^2}{\epsilon^2}\right)(\varepsilon\mu + \zeta^2) - \beta^2\right)H_x = 0$ <br> $E_z = \frac{1}{i\omega}\frac{\varepsilon}{\epsilon^2-\gamma^2}\left(\frac{\gamma\beta}{\varepsilon}H_x + \partial_y H_x - i\omega\zeta H_x\right)$ | 4.7 |
| **MO&ME effects** | Waveguide (TM mode) | $\begin{pmatrix}0 & 0 & 0\\0 & 0 & i\gamma\\0 & -i\gamma & 0\end{pmatrix}$ | $\begin{pmatrix}0 & 0 & 0\\0 & 0 & 0\\\zeta & 0 & 0\end{pmatrix}$ | $\partial_y\partial_y H_x + \left(\omega^2\varepsilon\mu\left(1-\frac{\gamma^2}{\epsilon^2}\right) + \omega^2\zeta^2 - \beta^2 - 2i\beta\frac{\omega\zeta\gamma}{\epsilon}\right)H_x = 0$ <br> $E_z = \left(\frac{\partial_y}{i\omega\varepsilon} - \frac{\zeta}{\varepsilon}\right)H_x$ | 4.7 |
| **MO&ME effects** | Waveguide (TM mode) | $\begin{pmatrix}0 & 0 & -i\gamma\\0 & 0 & 0\\i\gamma & 0 & 0\end{pmatrix}$ | $\begin{pmatrix}0 & 0 & 0\\0 & 0 & 0\\\zeta & 0 & 0\end{pmatrix}$ | $\partial_y^2 H_x + \left(\omega^2\left(\varepsilon\mu\left(1-\frac{\gamma^2}{\epsilon^2}\right)+\zeta^2\right) - \left(1-\frac{\gamma^2}{\epsilon^2}\right)\beta^2\right)H_x = 0$ <br> $E_z = \frac{\varepsilon}{\epsilon^2-\gamma^2}\left(\frac{\partial_y}{i\omega} - \zeta\right)H_x$ | 4.7 |
| **MO&ME effects** | Waveguide (TM mode) | $\begin{pmatrix}0 & i\gamma & 0\\-i\gamma & 0 & 0\\0 & 0 & 0\end{pmatrix}$ | $\begin{pmatrix}0 & 0 & 0\\\zeta & 0 & 0\\0 & 0 & 0\end{pmatrix}$ | $\partial_y^2\left(1-\frac{\gamma^2}{\epsilon^2}\right)H_x + \left(\omega^2\left(\mu\varepsilon\left(1-\frac{\gamma^2}{\epsilon^2}\right)+\zeta^2\right) - \beta^2\right)H_x = 0$ <br> $E_y = \frac{\varepsilon^2}{\epsilon^2-\gamma^2}\left(\frac{\beta}{\omega\varepsilon}-\frac{\zeta}{\varepsilon}\right)H_x$ | 4.7 |
| **MO&ME effects** | Waveguide (TM mode) | $\begin{pmatrix}0 & 0 & 0\\0 & 0 & i\gamma\\0 & -i\gamma & 0\end{pmatrix}$ | $\begin{pmatrix}0 & 0 & 0\\\zeta & 0 & 0\\0 & 0 & 0\end{pmatrix}$ | $\partial_y^2 H_x + \left(\omega^2\left(\mu\varepsilon\left(1-\frac{\gamma^2}{\epsilon^2}\right)+\zeta^2\right) - \beta^2\right)H_x + 2\partial_y\frac{\omega\zeta\gamma}{\epsilon}H_x = 0$ <br> $E_z = \frac{1}{i\omega}\frac{1}{\epsilon^2-\gamma^2}(\gamma(\beta-\omega\zeta)+\varepsilon\partial_y)H_x$ | 4.7 |
| **MO&ME effects** | Waveguide (TM mode) | $\begin{pmatrix}0 & 0 & -i\gamma\\0 & 0 & 0\\i\gamma & 0 & 0\end{pmatrix}$ | $\begin{pmatrix}0 & 0 & 0\\\zeta & 0 & 0\\0 & 0 & 0\end{pmatrix}$ | $\partial_y^2 H_x + \left(\omega^2\left(1-\frac{\gamma^2}{\epsilon^2}\right)(\mu\varepsilon+\zeta^2) - \left(1-\frac{\gamma^2}{\epsilon^2}\right)\beta^2\right)H_x = 0$ <br> $E_z = \frac{\partial_y}{i\omega}\frac{\varepsilon}{\epsilon^2-\gamma^2}H_x$ | 4.7 |

## 5. Analysis of cases with both MO and ME effects

In this section, a new model that accounts for both MO and ME effects is applied to free space and waveguides to discuss novel phenomena and device properties

*5.1 Nonreciprocal polarization conversion*

Based on the theory derived in Section 4.6, we discuss what properties can be obtained in free space propagation when MO and ME effects occur simultaneously.

For simplicity, the propagation properties of plane waves in a bulk medium are considered, as shown in Fig. 6. Here, we assume that MO and ME effects occur simultaneously in the medium and apply the same conditions under which we derived the wave equation in Section 4.6. In other words, we assume that the magnetization is aligned in the z direction ($a = \gamma, b = 0, c = 0$) and that the diagonal components G and H of the magneto-electric coupling tensor have values ($G = H = \zeta$).

The wave equation in the medium is expressed by Eq. (37). Assuming that the polarization state of the incident light is linearly polarized, the Faraday rotation angle of the forward wave $\theta_f$ and that of the backward wave $\theta_b$ in the medium are obtained as follows.

$$\theta_f = \left(\frac{\beta_{f-} - \beta_{f+}}{2}\right) z \qquad (42a)$$

$$\theta_b = \left(\frac{\beta_{b+} - \beta_{b-}}{2}\right) z \qquad (42b)$$

In addition, $\beta_{f\pm}$ and $\beta_{b\pm}$ are respectively the solution of the following equations.

$$f_\pm(\beta) = \beta^2 \mp i\omega\beta\zeta - \omega^2\varepsilon\mu - \omega^2\zeta^2 \pm \omega^2\mu\gamma = 0 \qquad (43a)$$

$$b_\pm(\beta) = \beta^2 \mp i\omega\beta\zeta - \omega^2\varepsilon\mu - \omega^2\zeta^2 \pm \omega^2\mu(-\gamma) = 0 \qquad (43b)$$

Here, the parameters $\gamma$, indicating MO effect, and $\zeta$, indicating ME effect, are both non-zero, which means that all four propagation constants $\beta_{f\pm}$ and $\beta_{b\pm}$ have different values. This makes it possible to control the Faraday rotation angle of the forward and backward waves almost independently by using two variables, $\gamma$ and $\zeta$, in Eq. (42).

We calculated the nonreciprocal polarization transformation for the configuration shown in Fig. 6, and the results are shown in Fig. 7. In this analysis, the permittivity $\varepsilon$ and the permeability $\mu$ of the bulk medium are set to 5 and 1, respectively. In addition, two parameters, the real part of the parameter $\gamma$ for MO effect and the imaginary part of the parameter $\zeta$ for ME effect, are taken as variables (the real part of $\gamma$ and the real part of $\zeta$ are set to 0). Figures 7(a) and (b) the effective refractive index difference between LCP and RCP is shown for the forward and backward waves, respectively.

$$\left|\frac{\beta_{f(b)+} - \beta_{f(b)-}}{k_0}\right| \qquad (44)$$

The horizontal axis shows the real part of $\gamma$, and the vertical axis shows the imaginary part of $\zeta$. The results show that the polarization rotation angles for the forward and backward waves can be determined separately by controlling two parameters. In particular, if we choose a parameter that sets the effective refractive index difference between LCP and RCP in the backward wave zero (white dotted line in Fig. 7), this suggests the possibility of realizing a unique nonreciprocal optical device in which the linear polarization rotates only in the forward wave.

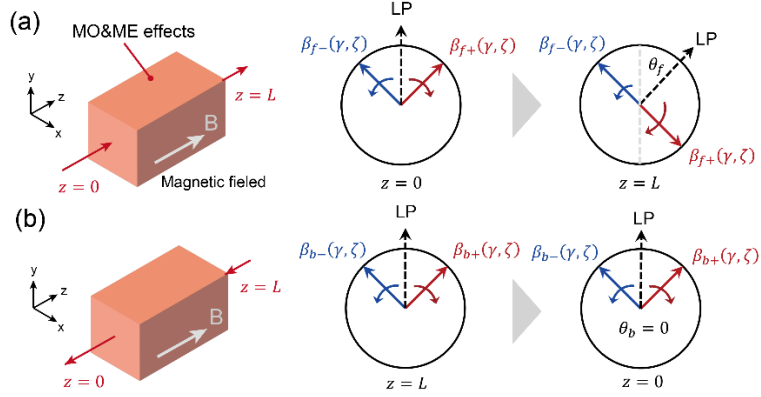

**Fig. 6.** Schematic diagram of polarization rotation in MO and ME effects in the case of (a) the forward wave and (b) the backward wave. (LP: Linear polarization)

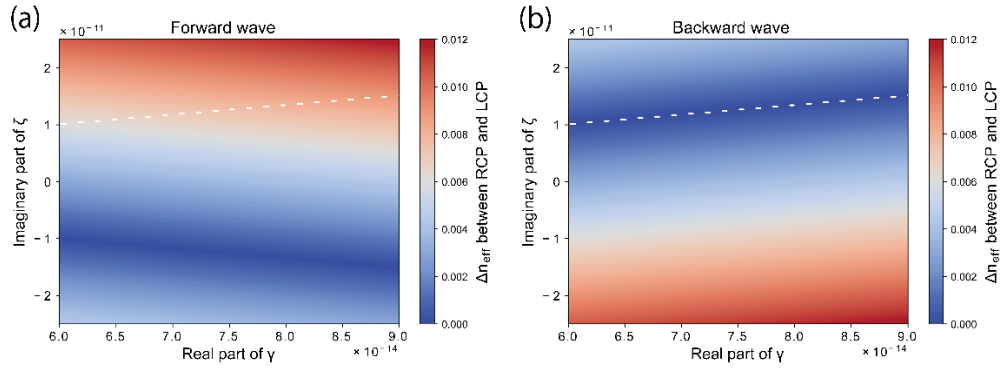

**Fig. 7.** Effective refractive index difference between LCP and RCP in (a) forward wave and (b) backward wave.

*5.2 Nonreciprocal effect amplification in waveguide*

In this section, we discuss what properties can be obtained in waveguide propagation where MO and ME effects occur simultaneously based on the theory derived in Section 4.7.

First, we consider TM mode propagation in a slab waveguide structure consisting of four layers, as shown in Fig. 8. It is assumed the top layer is the unique medium in which MO and ME effects occur simultaneously. In contrast, the other layers (core and cladding layers) consist of a general dielectric with isotropic properties. Here, the same conditions apply as in Section 4.7 for the top layer.

$$\begin{pmatrix} D_x \\ D_y \\ D_z \\ B_x \\ B_y \\ B_z \end{pmatrix} = \begin{pmatrix} \begin{pmatrix} \varepsilon_4 & 0 & 0 \\ 0 & \varepsilon_4 & i\gamma \\ c & -i\gamma & \varepsilon_4 \end{pmatrix} & \begin{pmatrix} 0 & 0 & 0 \\ 0 & 0 & 0 \\ \zeta & 0 & 0 \end{pmatrix} \\ -\begin{pmatrix} 0 & 0 & \zeta \\ 0 & 0 & 0 \\ 0 & 0 & 0 \end{pmatrix} & \begin{pmatrix} \mu & 0 & 0 \\ 0 & \mu & 0 \\ 0 & 0 & \mu \end{pmatrix} \end{pmatrix} \begin{pmatrix} E_x \\ E_y \\ E_z \\ H_x \\ H_y \\ H_z \end{pmatrix} \quad (45)$$

It is known that slab waveguides with MO effect cause nonreciprocal effects in forward and backward waves for TM mode light, as described in Section 4.3 [48–52]. In this study, we discuss how both MO and ME effects affect nonreciprocity compared to MO effect only.

In the slab waveguide shown in Fig. 8, the magnetic field $H_x$ and the electric field $E_z$ in the top and bottom layers are related by multiplying the transfer matrices in the middle layer.

$$\begin{pmatrix} H_{x4}(y_4) \\ \omega\varepsilon_0 E_{z4}(y_4) \end{pmatrix} = \mathbf{M_3} \cdot \mathbf{M_2} \cdot \begin{pmatrix} H_{x1}(y_1) \\ \omega\varepsilon_0 E_{z1}(y_1) \end{pmatrix} \tag{46}$$

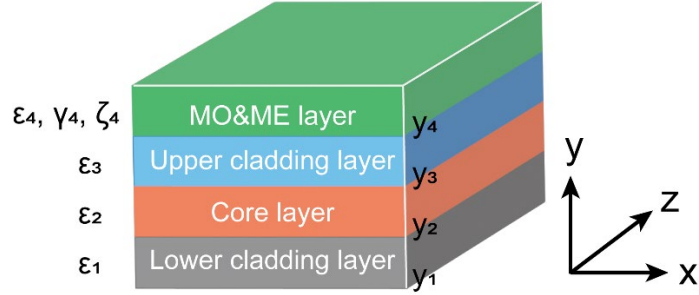

**Fig. 8.** Schematic cross section of our proposed waveguide

where $H_{xn}$, $E_{zn}$ are the magnetic field $H_x$ and the electric field $E_z$ of the $n$th layer, and $y_n$ is the bottom coordinate of the $n$th layer. The transfer matrix of the middle layer $\mathbf{M_2}$, $\mathbf{M_3}$ is given by the wave equation in Section 4.1 as follows.

$$\mathbf{M_{n\ (n=2\ or\ 3)}} = \begin{pmatrix} \cosh[\beta_n(y-y_n)] & \dfrac{i\varepsilon_n}{\beta_n}\sinh[\beta_n(y-y_n)] \\ \dfrac{\beta_n}{i\varepsilon_n}\sinh[\beta_n(y-y_n)] & \cosh[\beta_n(y-y_n)] \end{pmatrix} \tag{47a}$$

$$\beta_{n\ (n=2\ or\ 3)} = \sqrt{\beta^2 - k_0^2 \varepsilon_n} \tag{47b}$$

Under the assumption that the electromagnetic waves in the top ($n = 4$) and bottom ($n = 1$) layers are perfectly damped, then by the wave equations obtained in Section 4.7, $H_{x1}, E_{z1}$ and $H_{x4}, E_{z4}$ can each be expressed by the following equations, respectively.

$$H_{x1}(y) = \varGamma \exp[\beta_1(y-y_1)] \tag{48a}$$

$$E_{z1}(y) = \frac{1}{i\omega\varepsilon_0\varepsilon_1} \partial_y H_{x1}(y) \tag{48b}$$

$$H_{x4}(y) = \varLambda \exp[-\beta_4(y-y_4)] \tag{49a}$$

$$E_{z4}(y) = \frac{1}{i\omega\varepsilon_0} \frac{\varepsilon_4}{\varepsilon_4^2 - \gamma^2}\left(\frac{\gamma\beta}{\varepsilon_4} H_{x4}(y) + \partial_y H_{x4}(y) - i\omega\zeta H_{x4}(y)\right) \tag{49b}$$

where

$$\beta_1 = \sqrt{\beta^2 - k_0^2 \varepsilon_1} \tag{50a}$$

$$\beta_4 = \sqrt{\beta^2 - k_0^2 \varepsilon_4\left(1 - \frac{\gamma^2}{\varepsilon_4^2}\right) - \omega^2 \zeta^2 + i2\beta\frac{\omega\zeta\gamma}{\varepsilon_4}} \tag{50b}$$

After substituting Eqs. (48a), (48b), (49a), and (49b) into Eq. (46), let $y \to \infty$ on the left side of the equation and $y \to -\infty$ on the right side. Finally, after rearrangement on $\varGamma$ and $\varLambda$, we can derive the following equations.

$$\begin{pmatrix} m_{11} + m_{12}\dfrac{\beta_1}{i\varepsilon_1} & -1 \\ m_{21} + m_{22}\dfrac{\beta_1}{i\varepsilon_1} & \dfrac{\varepsilon_4}{i(\varepsilon_4^2 - \gamma^2)}\left(\beta_4 + i\omega\zeta - \dfrac{\gamma\beta_4}{\varepsilon_4}\right) \end{pmatrix} \begin{pmatrix} \varGamma \\ \varLambda \end{pmatrix} = \boldsymbol{R}\begin{pmatrix} \varGamma \\ \varLambda \end{pmatrix} = 0 \tag{51}$$

where $m_{ij}$ represents each element of the matrix $\mathbf{M_3} \cdot \mathbf{M_2}$. For Eq. (51) to have a non-trivial solution, the condition $\det R = 0$ must be satisfied. This allows us to determine the waveguide's propagation constants $\beta$ (refractive index and absorption coefficient). Since the first-order term of $\beta$ appears in Eq. (51) due to MO effect, the waveguide's propagation constants can be different for forward and backward waves by changing the sign of $\beta$. It also shows that their values vary with $\gamma$ due to MO effect.

Accordingly, Fig. 9 shows the calculation results of the nonreciprocal effects of the slab waveguide shown in Fig. 8. In this analysis, the dielectric constants $\varepsilon_2$, $\varepsilon_{1,3}$ and $\varepsilon_4$ of the core, upper and lower cladding layers, and top layer are set to $(3.49)^2$, $(3.16)^2$ and $(3.10)^2$, respectively. The parameter $\gamma$ related to MO effect on the top layer, was set to $-1 + i$. The real and imaginary parts of the parameter $\zeta$ related to ME effect, were used as variables to estimate the nonreciprocal effects (see Table 4).

Fig. 9(a) and Fig. 9(b) plot the analytical results of the NRL and NRPS respectively, where the horizontal axis represents the real part of $\zeta$ and the vertical axis represents the imaginary part of $\zeta$. The NRL is a parameter used to quantitatively discuss the loss difference between the forward and backward waves, and is given by the following equation [53].

$$\text{NRL[dB/mm]} = \frac{10}{\ln 10}\text{NRL[1/mm]} = \left| \frac{2\text{Im}(\beta_{forwad})}{\text{Re}\left(n_{eff_{forward}}\right)} - \frac{2\text{Im}(\beta_{backwad})}{\text{Re}\left(n_{eff_{backward}}\right)} \right| \times 10^{-3} \quad (52)$$

The NRPS is also used to quantitatively discuss the phase difference between the forward and backward waves during propagation. It is given by the following equation [54].

$$\text{NRPS[rad/mm]} = \left|\text{Re}(\beta_{forwad}) - \text{Re}(\beta_{backward})\right| \times 10^{-3} \quad (53)$$

The origin of the graph in Fig. 9 is identical to that of the waveguide optical isolator using the transverse Kerr effect (nonreciprocal effect in a slab waveguide with a medium having only MO effect as the top layer) since the real and imaginary parts of $\zeta$ are both zero. These results suggest that the interaction of MO and ME effects may significantly improve the nonreciprocity compared to MO effect alone.

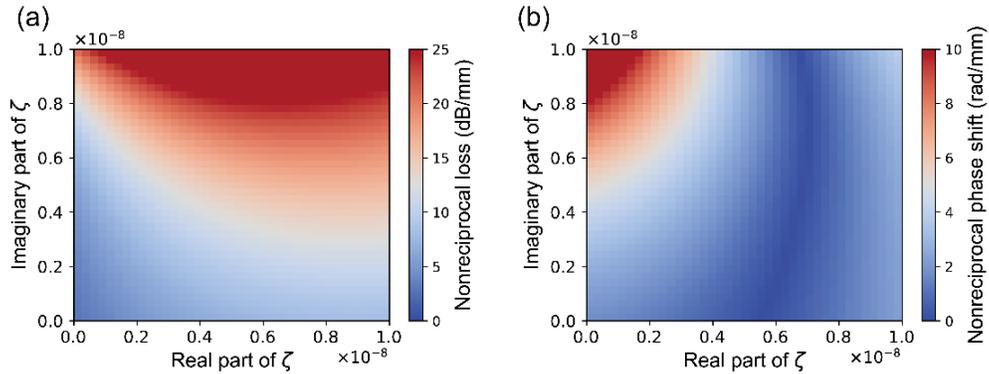

**Fig. 9.** Nonreciprocal loss (a) and phase shift (b) for MO and ME effects.

Table 4 Parameters used for calculating device characteristics

| Parameters | Values | Layers |
|---|---|---|
| $\varepsilon_1$ | $(3.16)^2$ | Lower cladding layer |
| $\varepsilon_2$ | $(3.49)^2$ | Core layer |
| $\varepsilon_3$ | $(3.16)^2$ | Upper cladding layer |
| $\varepsilon_4$ | $(3.10 + 3i)^2$ | MO&ME layer |
| $\gamma_4$ | $-1 + i$ | MO&ME layer |

## 6. Conclusions

In this study, an optical waveguide theory that can comprehensively discuss MO and ME effects is proposed. This theory allows propagation properties to be analyzed for all medium that exhibit MO and ME effects. An optical waveguide containing metamaterials and ferromagnetic materials is also proposed to control MO and ME effects. It has been shown that the interaction between MO and ME effects can come about through the choice of the arrangement of the metamaterials and the direction of magnetization. Thus, it has been shown by simulations that a polarization control function, such as a polarization rotation in only one direction, is possible in free-space propagation. In addition, it is possible to improve the nonreciprocal nature of propagating light for waveguide propagation. If the ME parameters shown in this study can be realized by using multiferroic materials and metamaterials, it is expected that entirely new polarization control devices and optical isolators with smaller size and higher performance will be possible.

**Disclosures.** The authors declare no conflicts of interest.

**Data availability.** Data underlying the results presented in this paper are not publicly available at this time but may be obtained from the authors upon reasonable request.